\documentclass[aps,pra,twocolumn,nofootinbib]{revtex4}%
\usepackage{graphics}
\usepackage{amsmath}
\usepackage{graphicx}
\usepackage{amsfonts}
\usepackage{amssymb}
\usepackage{float}
\usepackage{longtable}
\usepackage{epsfig}
\usepackage{latexsym}
\usepackage{theorem}
\usepackage{bbm}
\usepackage{bm}
\usepackage{psfrag}
\newtheorem{theorem}{Theorem}
\newtheorem{lemma}[theorem]{Lemma}

\begin{document}
\title{Ultimate precision of adaptive noise estimation}
\author{Stefano Pirandola}
\author{Cosmo Lupo}
\affiliation{Computer Science \& York Centre for Quantum Technologies, University of York,
York YO10 5GH, UK}

\begin{abstract}
We consider the estimation of noise parameters in a quantum channel, assuming
the most general strategy allowed by quantum mechanics. This is based on the
exploitation of unlimited entanglement and arbitrary quantum operations, so
that the channel inputs may be interactively updated. In this general scenario
we draw a novel connection between quantum metrology and teleportation. In
fact, for any teleportation-covariant channel (e.g., Pauli, erasure, or
Gaussian channel), we find \ that adaptive noise estimation cannot beat the
standard quantum limit, with the quantum Fisher information being determined
by the channel's Choi matrix. As an example, we establish the ultimate
precision for estimating excess noise in a thermal-loss channel which is
crucial for quantum cryptography. Because our general methodology applies to
any functional which is monotonic under trace-preserving maps, it can be
applied to simplify other adaptive protocols, including those for quantum
channel discrimination. Setting the ultimate limits for noise estimation and
discrimination paves the way for exploring the boundaries of quantum sensing,
imaging and tomography.

\end{abstract}
\maketitle



Quantum metrology~\cite{Sam1,Sam2,Paris,Giova,ReviewNEW} deals with the
optimal estimation of classical parameters encoded in quantum transformations.
Its applications are many, from enhancing gravitational wave
detectors~\cite{grav1,grav2}, to improving frequency standards~\cite{freq},
clock synchronization~\cite{clock2} and optical
resolution~\cite{Lupo16,Tsang15,Tsang2}, just to name a few. Understanding its
ultimate limits is therefore of paramount importance. However, it is also
challenging, because the most general strategies for quantum parameter
estimation exploit adaptive, i.e., feedback-assisted, quantum operations (QOs)
involving an arbitrary number of ancillas.

Adaptive protocols are difficult to
study~\cite{adaptive1,adaptive2,ada3,ada4,ada4bis,ada5} but a powerful tool
can now be borrowed from the field of quantum communication. In this context,
Ref.~\cite{PLOB} has recently designed a general and dimension-independent
technique which reduces adaptive protocols into a block form. This technique
of \textquotedblleft teleportation stretching\textquotedblright\ is
particularly powerful when the protocols are implemented over suitable
teleportation-covariant channels~\cite{PLOB}, which are those channels
commuting with the random unitaries induced by teleportation. This is a broad
class, including Pauli, erasure~\cite{Nielsen}, and bosonic Gaussian
channels~\cite{WeeRMP}.

In this work, we exploit the tool of teleportation stretching to simplify
adaptive protocols of quantum metrology. We discover that the adaptive
estimation of noise in a teleportation-covariant channel cannot beat the
standard quantum limit (SQL). Our no-go theorem also establishes that this
limit is achievable by using entanglement without adaptiveness, so that the
quantum Fisher information (QFI)~\cite{Sam1} assumes a remarkably simple
expression in terms of the channel's Choi matrix. As an application, we set
the ultimate adaptive limit for estimating thermal noise in Gaussian channels,
which has implications for continuous-variable quantum key distribution (QKD)
and, more generally, for measurements of temperature in quasi-monochromatic
bosonic baths.

Because our methodology applies to any functional of quantum states which is
monotonic under completely-positive trace-preserving (CPTP) maps, we may
simplify other types of adaptive protocols, including those for quantum
hypothesis testing~\cite{QHT,QHT2,Invernizzi,QHB1,Gae1}. Here we find that the
ultimate error probability for discriminating two teleportation-covariant
channels is reached without adaptiveness and determined by their Choi
matrices. Applications are for protocols of quantum sensing, such as quantum
reading~\cite{Qread,Qread2,Qread3,Qread4,Nair11,Hirota11,Bisio11,Arno11} and
illumination~\cite{Qill0,Qill1,Qill2,Qill3}, and for the resolution of
extremely-close temperatures~\cite{field1,field2}.

\bigskip

\textit{Adaptive protocols for quantum parameter estimation.}--~~The most
general adaptive protocol for quantum parameter estimation can be formulated
as follows. Let us consider a box containing a quantum channel $\mathcal{E}%
_{\theta}$ characterized by an unknown classical parameter $\theta$. We then
pass this box to Alice and Bob, whose task is to retrieve the best estimate of
$\theta$. Alice prepares the input to probe the box, while Bob gets the
corresponding output. The parties may exploit unlimited entanglement and apply
joint QOs before and after each probing.
These QOs may distribute entanglement and contain measurements that can always
be post-poned at the end of the protocol (thanks to the principle of deferred
measurement~\cite{Nielsen}).

In our formulation, we assume that Alice has a local register with an ensemble
of systems $\mathbf{a}=\{a_{1},a_{2},...\}$. Similarly, Bob has another local
register $\mathbf{b}=\{b_{1},b_{2},...\}$. These registers are intended to be
dynamic, so that they can be depleted or augmented with quantum systems. Thus,
when Alice picks an input system $a\in\mathbf{a}$, we update her register as
$\mathbf{a\rightarrow a}a$. Then, suppose that system $a$ is transmitted to
Bob, who receives the output system $b$. The latter is stored in his register,
updated as $\mathbf{b}b\mathbf{\rightarrow b}$.

\begin{figure}[ptbh]
\vspace{-1.2cm}
\par
\begin{center}
\vspace{-0.7cm} \includegraphics[width=0.53\textwidth]{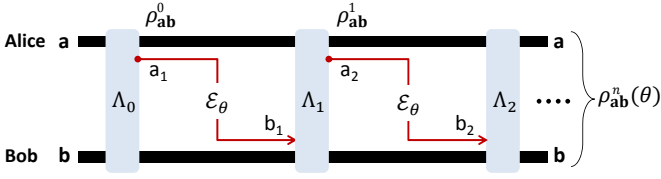}
\vspace{-0.8cm}
\end{center}
\par
\vspace{-2.2cm}\caption{Arbitrary adaptive protocol for quantum parameter
estimation. After preparation of the register state $\rho_{\mathbf{ab}}^{0}$
by means of an initial QO $\Lambda_{0}$, Alice starts probing the box
$\{\mathcal{E}_{\theta}\}$ by sending a system $a_{1}$ from her register, with
Bob getting the output $b_{1}$. This is repeated$\ n$ times with each
transmission $a_{i}\rightarrow b_{i}$ interleaved by two QOs $\Lambda_{i-1}$
and $\Lambda_{i}$. The output state $\rho_{\mathbf{ab}}^{n}(\theta)$ is
finally subject to an optimal measurement.}%
\label{probingFIG}%
\end{figure}

The first part of the protocol is the preparation of the initial register
state $\rho_{\mathbf{ab}}^{0}$ by applying the first QO $\Lambda_{0}$ to some
fundamental state. After this preparation, the parties start the adaptive
probings. Alice picks a system $a_{1}\in\mathbf{a}$ and send it through the
box $\{\mathcal{E}_{\theta}\}$. At the output, Bob receives a\ system $b_{1}$,
which is stored in his register $\mathbf{b}$. At the end of the first probing,
the two parties applies a joint QO $\Lambda_{1}$, which updates and optimizes
their registers for the next uses. In the second probing, Alice picks another
system $a_{2}\in\mathbf{a}$, sends it through the box, with Bob receiving
$b_{2}$ and so on. After $n$ probings, we have a sequence of QOs
$\mathcal{P}=\{\Lambda_{0},\ldots,\Lambda_{n}\}$ generating an output state
$\rho_{\mathbf{ab}}^{n}(\theta)$ for Alice and Bob~\cite{averageLOCC}. See
Fig.~\ref{probingFIG}.

The final step consists of measuring the output state. The outcome is
processed into an unbiased estimator of $\theta$, with an associated
protocol-dependent QFI%
\begin{equation}
I_{\theta}^{n}(\mathcal{P})=\frac{8\left\{  1-F[\rho_{\mathbf{ab}}^{n}%
(\theta),\rho_{\mathbf{ab}}^{n}(\theta+d\theta)]\right\}  }{d\theta^{2}},
\label{formSAM}%
\end{equation}
with $F(\rho,\sigma):=\mathrm{Tr}\sqrt{\sqrt{\sigma}\rho\sqrt{\sigma}}$ being
the fidelity~\cite{BuresFID}. By optimizing over all adaptive protocols, we
define the adaptive QFI $\bar{I}_{\theta}^{n}:=\sup_{\mathcal{P}}I_{\theta
}^{n}(\mathcal{P})$, so that the minimum error-variance in the estimation of
$\theta$ satisfies the quantum Cramer-Rao bound (QCRB)~\cite{Sam1,Sam2}
$\mathrm{Var}(\theta)\geq1/\bar{I}_{\theta}^{n}$.

\bigskip

\textit{Teleportation stretching for quantum metrology.}--~We now compute the
adaptive QFI. Consider the class of teleportation-covariant channels in
arbitrary dimension as generally defined in Ref.~\cite{PLOB}. They correspond
to those quantum channels commuting with the random unitaries induced by
teleportation, which are Pauli operators at finite dimension and displacement
operators at infinite dimension~\cite{telereview,teleBennett,Samtele}. By
definition, a quantum channel $\mathcal{E}$ is called \textquotedblleft
teleportation-covariant\textquotedblright\ if, for any teleportation unitary
$U$ we may write~\cite{PLOB}%
\begin{equation}
\mathcal{E}(U\rho U^{\dagger})=V\mathcal{E}(\rho)V^{\dagger},
\label{stretchability}%
\end{equation}
for some other unitary $V$. This is a common property, owned by Pauli,
erasure, and bosonic Gaussian channels.

\begin{figure*}[ptbh] \vspace{-1.2cm}
\par
\begin{center}
\vspace{-1.9cm} \includegraphics[width=0.73\textwidth]{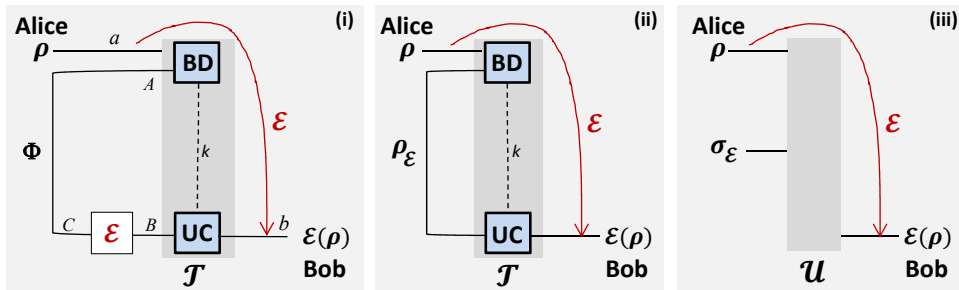}
\vspace{+0.4cm} \vspace{-2.3cm}
\end{center}
\par
\vspace{-1.2cm}\caption{Teleportation covariance and channel simulation. In
panel~\textbf{(i)}, we consider a teleportation-covariant channel
$\mathcal{E}$ (red curvy line) from Alice's system $a$ to Bob's system\ $b$.
This can be simulated by teleporting system $a$ to system $C$, by means of a
maximally-entangled state $\Phi_{AC}$ and a Bell detection (BD) on systems $a$
and $A$, with outcome $k$. System $C$ is projected onto a state $\rho_{C}$
which is equal to $\rho_{a}$ up to a teleportation unitary $U_{k}$. Because of
Eq.~(\ref{stretchability}), we now have $\rho_{B}=\mathcal{E}(\rho
_{C})=\mathcal{E}(U_{k}\rho_{a}U_{k}^{\dagger})=V_{k}\mathcal{E}(\rho
_{a})V_{k}^{\dagger}$ for some other unitary $V_{k}$. Upon receiving $k$ from
Alice, Bob may undo $V_{k}$ on system $B$ by applying a unitary correction
(UC) $V_{k}^{-1}$. Thus, he retrieves the output state $\rho_{b}%
=\mathcal{E}(\rho_{a})$. Overall, Alice's BD and Bob's UC represent a
teleportation LOCC $\mathcal{T}$. As shown in panel~\textbf{(ii)}, this is
equivalent to simulate the channel by teleporting \ the state over the
channel's Choi matrix $\rho_{\mathcal{E}}:=\mathcal{I}\otimes\mathcal{E}%
(\Phi)$, so that we may write Eq.~(\ref{teleSIM}). The teleportation
simulation $(\mathcal{T},\rho_{\mathcal{E}})$ becomes asymptotic
$(\mathcal{T}_{\mu},\rho_{\mathcal{E}}^{\mu})$ for bosonic channels. By
comparing with panel~\textbf{(iii)}, we see that we have provided a computable
design for the tool of quantum simulation~\cite{Gatearray,Qsim0,Qsim},
reducing the quantum operation $\mathcal{U}$ to a teleportation LOCC
$\mathcal{T}$, and the (difficult-to-find) programme state $\sigma
_{\mathcal{E}}$ to the channel's Choi matrix $\rho_{\mathcal{E}}$.}%
\label{trio}%
\end{figure*}

Because of Eq.~(\ref{stretchability}), we can simulate the channel
$\mathcal{E}$\ via local operations and classical communication (LOCC) applied
to a suitable resource state. In fact, as explained in Fig.~\ref{trio}(i-ii),
channel $\mathcal{E}$ can be simulated by a teleportation LOCC $\mathcal{T}$
performed over the channel's Choi matrix $\rho_{\mathcal{E}}$, i.e., we may
write~\cite{PLOB}
\begin{equation}
\mathcal{E}(\rho)=\mathcal{T}(\rho\otimes\rho_{\mathcal{E}}). \label{teleSIM}%
\end{equation}
This simulation is intended to be asymptotic for bosonic
channels~\cite{PLOB}. We consider
$\mathcal{E}(\rho)=\lim_{\mu}\mathcal{T}_{\mu}(\rho\otimes
\rho_{\mathcal{E}}^{\mu})$, where $\mathcal{T}_{\mu}$ is a
sequence of teleportation LOCCs and
$\rho_{\mathcal{E}}^{\mu}:=\mathcal{I}\otimes
\mathcal{E}(\Phi^{\mu})$ is a sequence computed on two-mode
squeezed vacuum (TMSV) states $\Phi^{\mu}$~\cite{WeeRMP}, so that
$\Phi:=\lim_{\mu}\Phi^{\mu}$ defines the asymptotic
Einstein-Podolsky-Rosen (EPR) state and $\rho
_{\mathcal{E}}:=\lim_{\mu}\rho_{\mathcal{E}}^{\mu}$ defines the
asymptotic Choi matrix~\cite{PLOB}. In the following, for any pair
of asymptotic states $\rho_{0,1}:=\lim_{\mu}\rho_{0,1}^{\mu}$, we
correspondingly extend a functional $f$ to the limit as
$f(\rho_{0},\rho_{1}):=\lim_{\mu}f(\rho
_{0}^{\mu},\rho_{1}^{\mu})$. 

The teleportation-based simulation provides a powerful design to the generic
tool of quantum simulation~\cite{Gatearray,Qsim0,Qsim} which is described by
\begin{equation}
\mathcal{E}(\rho)=\mathcal{U}(\rho\otimes\sigma_{\mathcal{E}}), \label{NCppp}%
\end{equation}
where $\mathcal{U}$ is a trace-preserving QO~\cite{parameterU} and
$\sigma_{\mathcal{E}}$ is some programme state, as in Fig.~\ref{trio}(iii).
First of all, we establish a simple criterion (teleportation covariance) that
allows us to identify channels $\mathcal{E}$ that are simulable as in
Eq.~(\ref{teleSIM}) and, therefore, programmable as in Eq.~(\ref{NCppp}).
Then, we give an explicit solution to Eq.~(\ref{NCppp}), so that $\mathcal{U}$
reduces to teleportation and the programme state $\sigma_{\mathcal{E}}$ is
found to be the channel's Choi matrix (see Fig.~\ref{trio}). As we will see
below, this insight drastically simplifies computations.

For a channel which is \textquotedblleft Choi-stretchable\textquotedblright%
\ as in Eq.~(\ref{teleSIM}), we may apply teleportation
stretching~\cite{PLOB,Network}. After stretching, the output $\rho
_{\mathbf{ab}}^{n}$ of an adaptive protocol for quantum/private communication
takes the form
\begin{equation}
\rho_{\mathbf{ab}}^{n}=\bar{\Lambda}(\rho_{\mathcal{E}}^{\otimes n}),
\label{stretchMAIN}%
\end{equation}
where $\bar{\Lambda}$ is trace-preserving LOCC~\cite{Note}. Here, to simplify
quantum metrology, we do not need to enforce the LOCC structure, so that
$\bar{\Lambda}$ may be an arbitrary CPTP map. In this sense the following
lemma provides a full adaptation of the tool for the task of parameter
estimation~\cite{Appendix}.

\begin{lemma}
[stretching of adaptive metrology]\label{LemmaSTRETCH}Consider the adaptive
estimation of the parameter $\theta$ of a teleportation-covariant channel
$\mathcal{E}_{\theta}$. After $n$ probings, the output of the adaptive
protocol can be written as%
\begin{equation}
\rho_{\mathbf{ab}}^{n}(\theta)=\bar{\Lambda}\left(  \rho_{\mathcal{E}_{\theta
}}^{\otimes n}\right)  =\lim_{\mu}\bar{\Lambda}_{\mu}(\rho_{\mathcal{E}%
_{\theta}}^{\mu\otimes n}), \label{ChoiDEC}%
\end{equation}
where $\bar{\Lambda}$ is a $\theta$-independent CPTP map and $\rho
_{\mathcal{E}_{\theta}}$ is the channel's Choi matrix. If channel
$\mathcal{E}_{\theta}$ is bosonic, then the decomposition is asymptotic
$(\bar{\Lambda}_{\mu},\rho_{\mathcal{E}_{\theta}}^{\mu})$ with a sequence of
CPTP maps $\bar{\Lambda}_{\mu}$ and Choi-approximating states $\rho
_{\mathcal{E}_{\theta}}^{\mu}$.
\end{lemma}

By exploiting Lemma~\ref{LemmaSTRETCH}, we now show that the adaptive
estimation of noise in teleportation-covariant channels cannot exceed the SQL,
and can always be reduced to non-adaptive strategies. In fact, we have the
following no-go theorem from teleportation~\cite{Appendix}.

\begin{theorem}
[No-go: tele-covariance implies SQL]\label{theoSTT}The adaptive estimation of
the noise parameter $\theta$ of a teleportation-covariant channel
$\mathcal{E}_{\theta}$ satisfies the QCRB $\mathrm{Var}(\theta)\geq1/\bar
{I}_{\theta}^{n}$, where the adaptive QFI takes the form%
\begin{equation}
\bar{I}_{\theta}^{n}=nB(\rho_{\mathcal{E}_{\theta}}),~~B(\rho_{\mathcal{E}%
_{\theta}}):=\frac{8\left[  1-F(\rho_{\mathcal{E}_{\theta}},\rho
_{\mathcal{E}_{\theta+d\theta}})\right]  }{d\theta^{2}}. \label{firstTH}%
\end{equation}
For large $n$, the QCRB is achievable by entanglement-based non-adaptive
protocols. For bosonic channels, we implicitly assume $F(\rho_{\mathcal{E}%
_{\theta}},\rho_{\mathcal{E}_{\theta+d\theta}}):=\lim_{\mu}F(\rho
_{\mathcal{E}_{\theta}}^{\mu},\rho_{\mathcal{E}_{\theta+d\theta}}^{\mu})$.
\end{theorem}

There are two important aspects in this theorem. The first is the
achievability of the bound~\cite{Achievability}. The second is the extreme
simplification of the adaptive QFI, which becomes a functional of the
channel's Choi matrix, computable almost instantaneously for many channels.
Because the QFI\ takes such a simple form, our results are easily extended to
bosonic channels~\cite{lossnoise} and can also be generalized to
multiparameter estimation~\cite{Appendix}. The teleportation-based approach is
so powerful that it is an open problem to find other channels (e.g.,
programmable) for which we may compute the adaptive QFI beyond the class of
teleportation-covariant channels.


\bigskip

\textit{Analytical formulas.}--~~Let us use Theorem~\ref{theoSTT} to study the
adaptive estimation of error probabilities in qubit channels~\cite{Nielsen}.
For a depolarizing channel with probability $p$ we find the asymptotically
achievable bound~\cite{Appendix}
\begin{equation}
\mathrm{Var}(p)\geq p(1-p)/n. \label{res2}%
\end{equation}
This result is also valid for the adaptive estimation of the probability $p$
of a dephasing channel or an erasure channel~\cite{Appendix}. Thus we show
that the bounds of Refs.~\cite{Qsim,RafaComms} are adaptive in a
straightforward way.

Now consider a bosonic Gaussian channel which transforms input
quadratures~\cite{WeeRMP} $\hat{x}=(\hat{q},\hat{p})^{T}$ as $\hat
{x}\rightarrow\eta\hat{x}+|1-\eta|\hat{x}_{T}+\xi$, where $\eta$ is a real
gain parameter, $\hat{x}_{T}$ are the quadratures of a thermal environment
with $\bar{n}_{T}$ mean number of photons, and $\xi$ is an additive Gaussian
noise variable with variance $w$. A specific case is the thermal-loss channel
for which $0\leq\eta<1$ and $\xi=0$. It is immediate to compute the ultimate
(adaptive) limit for estimating thermal noise $\bar{n}_{T}>0$ in such a
channel. By using our Theorem~\ref{theoSTT} and the formula for the fidelity
between multimode Gaussian states~\cite{Banchi}, we easily
derive~\cite{Appendix}%
\begin{equation}
\mathrm{Var}(\bar{n}_{T})\geq\bar{n}_{T}(\bar{n}_{T}+1)/n, \label{varttt}%
\end{equation}
which is achievable for large $n$.

The latter result sets the ultimate precision for estimating the excess
(thermal) noise in a tapped communication line~\cite{QKD} or the temperature
of a quasi-monochromatic bosonic bath. Eq.~(\ref{varttt}) is also valid for
estimating thermal noise in an amplifier, defined by $\eta>1$ and $\xi=0$.
Finally, for $\eta=1$ and $\xi\neq0$, we have an additive-noise Gaussian
channel. The adaptive estimation of its variance $w>0$ is limited
by~\cite{Appendix}
\begin{equation}
\mathrm{Var}(w)\geq w^{2}/n.
\end{equation}

\bigskip

\textit{Adaptive quantum channel discrimination.}--~~We can simplify other
types of adaptive protocols whose performance is quantified by functionals
which are monotonic under CPTP maps~\cite{MethodsNOTA}. Thus, consider a box
with two equiprobable channels $\{\mathcal{E}_{k}\}=\{\mathcal{E}%
_{0},\mathcal{E}_{1}\}$. An adaptive discrimination protocol $\mathcal{P}$
consists of local registers prepared in a state $\rho_{\mathbf{ab}}^{0}$,
which are then used to probe the box $n$ times while being assisted by a
sequence of QOs $\mathcal{P}$, similar to Fig.~\ref{probingFIG}. The output
state $\rho_{\mathbf{ab}}^{n}(k)$ is optimally measured~\cite{Hesltrom} so
that we may write the protocol-dependent error probability in terms of the
trace distance $D$%
\begin{equation}
p(k^{\prime}\neq k|\mathcal{P})=\frac{1-D[\rho_{\mathbf{ab}}^{n}%
(0),\rho_{\mathbf{ab}}^{n}(1)]}{2}. \label{discrPP}%
\end{equation}
The ultimate error probability is given by optimizing over all adaptive
protocols, i.e., $p_{\mathrm{err}}:=\inf_{\mathcal{P}}p(k^{\prime}\neq
k|\mathcal{P})$.

For the discrimination of teleportation-covariant channels, we may write the
output state $\rho_{\mathbf{ab}}^{n}(k)$\ using the same Choi decomposition of
Eq.~(\ref{ChoiDEC}), proviso that we replace $\rho_{\mathcal{E}_{\theta}}$
with its discrete version $\rho_{\mathcal{E}_{k}}$, i.e.,
\begin{equation}
\rho_{\mathbf{ab}}^{n}(k)=\bar{\Lambda}\left(  \rho_{\mathcal{E}_{k}}^{\otimes
n}\right)  , \label{discrSS}%
\end{equation}
understood to be asymptotic for bosonic channels. We then
prove~\cite{Appendix} the following result which expresses $p_{\mathrm{err}}$
in terms of the trace distance between Choi matrices.

\begin{theorem}
\label{theoDIS}Consider an adaptive protocol for discriminating two
teleportation-covariant channels $\{\mathcal{E}_{0},\mathcal{E}_{1}\}$. After
$n$ probings, the minimum error probability is%
\begin{equation}
p_{\mathrm{err}}=\frac{1-D(\rho_{\mathcal{E}_{0}}^{\otimes n},\rho
_{\mathcal{E}_{1}}^{\otimes n})}{2}, \label{discrTHEO}%
\end{equation}
where $D=\lim_{\mu}D[\rho_{\mathcal{E}_{0}}^{\mu\otimes n},\rho_{\mathcal{E}%
_{1}}^{\mu\otimes n}]$ for bosonic channels.
\end{theorem}

For programmable channels $\{\mathcal{E}_{k}\}$ with states $\{\sigma
_{\mathcal{E}_{k}}\}$, we may only write the bound $p_{\mathrm{err}}%
\geq\lbrack1-D(\sigma_{\mathcal{E}_{0}}^{\otimes n},\sigma_{\mathcal{E}_{1}%
}^{\otimes n})]/2$. In general, this is not achievable because we do not know
if $\sigma_{\mathcal{E}_{k}}$ can be generated by transmission through
$\mathcal{E}_{k}$. By contrast, for teleportation-covariant channels, the
bound is always achievable and the optimal strategy is non-adaptive, based on
sending parts of maximally-entangled states and then measuring the output Choi
matrices. Because of the equality in Eq.~(\ref{discrTHEO}) we may write both
lower and upper (single-letter) bounds. Using the Fuchs-van der Graaf
relations~\cite{Fuchs}, the quantum Pinsker's inequality~\cite{Pinsker,Lieb},
and the quantum Chernoff bound (QCB)~\cite{QCB1}, we find that the adaptive
discrimination of teleportation-covariant channels must
satisfy~\cite{Appendix}
\begin{equation}
\frac{1-\sqrt{\min\left\{  1-F^{2n},nS\right\}  }}{2}\leq p_{\mathrm{err}}%
\leq\frac{Q^{n}}{2}\leq\frac{F^{n}}{2}, \label{REbound}%
\end{equation}
where $F:=F(\rho_{\mathcal{E}_{0}},\rho_{\mathcal{E}_{1}})$, $Q:=\inf
_{s}\mathrm{Tr}(\rho_{\mathcal{E}_{0}}^{s},\rho_{\mathcal{E}_{1}}^{1-s})$ and
$S:=(\ln\sqrt{2})\min\{S(\rho_{\mathcal{E}_{0}}||\rho_{\mathcal{E}_{1}%
}),S(\rho_{\mathcal{E}_{1}}||\rho_{\mathcal{E}_{0}})\}$, with $S(\rho
||\sigma)$ being the relative entropy~\cite{VedralRMP}. Here recall that the
QCB is tight for large $n$~\cite{QCB1}, so that $p_{\mathrm{err}}\simeq
Q^{n}/2$. All these functionals are asymptotic for bosonic channels.

In particular, for two thermal-loss channels with identical transmissivity but
different thermal noise, $\bar{n}_{0}$ and $\bar{n}_{1}$, we may take the
limit and compute~\cite{Appendix}%
\begin{equation}
Q=\inf_{s}\left[  (\bar{n}_{0}+1)^{s}(\bar{n}_{1}+1)^{1-s}-\bar{n}_{0}^{s}%
\bar{n}_{1}^{1-s}\right]  ^{-1}.
\end{equation}
For these channels, it is interesting to study the infinitesimal
discrimination $\bar{n}_{0}=\bar{n}_{T}$ and $\bar{n}_{1}=\bar{n}_{T}+d\bar
{n}_{T}$. As we show in a lemma~\cite{Appendix}, when we consider the
discrimination of two infinitesimally-close states, $\rho_{\theta}$ and
$\rho_{\theta+d\theta}$, the $n$-copy minimum error probability can be
connected with the QCRB for estimating parameter $\theta$. Applying this
result to the asymptotic Choi matrices of the thermal-loss channels and taking
the limit of large $n$, we get~\cite{Appendix} $p_{\mathrm{err}}\simeq
e^{-n\Sigma}/2$ where $\Sigma=[8\bar{n}_{T}(\bar{n}_{T}+1)]^{-1}d\bar{n}%
_{T}^{2}$ for $\bar{n}_{T}>0$. For the specific case of $\bar{n}_{T}=0$
(infinitesimal discrimination from vacuum noise), we have a discontinuity and
we may write $\Sigma=d\bar{n}_{T}$~\cite{Appendix}. These results represent
the ultimate adaptive limits for resolving two temperatures, e.g., for testing
the Unruh effect~\cite{field1} or the Hawking radiation in analogue
systems~\cite{field2}.


\bigskip

\textit{Conclusions}.--~~In this paper we have established the ultimate limits
of adaptive noise estimation and discrimination for the wide class of
teleportation-covariant channels, which includes fundamental transformations
for qubits, qudits and bosonic systems. We have reduced the most general
adaptive protocols for parameter estimation and channel discrimination into
much simpler block versions, where the output states are simply expressed in
terms of Choi matrices of the encoding channels. This allowed us to prove that
the optimal noise estimation of teleportation-covariant channels scales as the
SQL and is fully determined by their Choi matrices. Our work not only shows
that teleportation is a primitive for quantum metrology but also provides
remarkably simple and practical results, such as the precision limit for
estimating the excess noise of a thermal-loss channel, which is a basic
channel in continuous variable QKD. Setting the ultimate precision limits of
noise estimation and discrimination has broad implications, e.g., in quantum
tomography, imaging, sensing and even for testing quantum field theories in
non-inertial frames.

\smallskip

\textit{Acknowledgments}.--~~This work was supported by the UK
Quantum Communications hub (EP/M013472/1) and the Innovation Fund
Denmark (Qubiz project). The authors thank S. Lloyd, S. L.
Braunstein, R. Laurenza, L. Maccone, R. Demkowicz-Dobrzanski, D.
Braun, and J. Kolodynski for comments and discussions.

\smallskip

\textit{Note added}.--~~While completing the final revision of this work, a
follow-up~\cite{Copy} appeared on the arXiv~\cite{AddedNOTE}.




\newpage\onecolumngrid

\newcounter{S} \setcounter{section}{0} \setcounter{subsection}{0}
\renewcommand{\bibnumfmt}[1]{[S#1]} {} \renewcommand{\citenumfont}[1]{S#1}

\section*{Supplemental Material}

\section{Teleportation stretching of adaptive quantum metrology (proof of
Lemma~1)\label{app1}}

\begin{figure}[tbh]
\vspace{-0.4cm}
\par
\begin{center}
\vspace{-4cm} \includegraphics[width=0.98\textwidth]{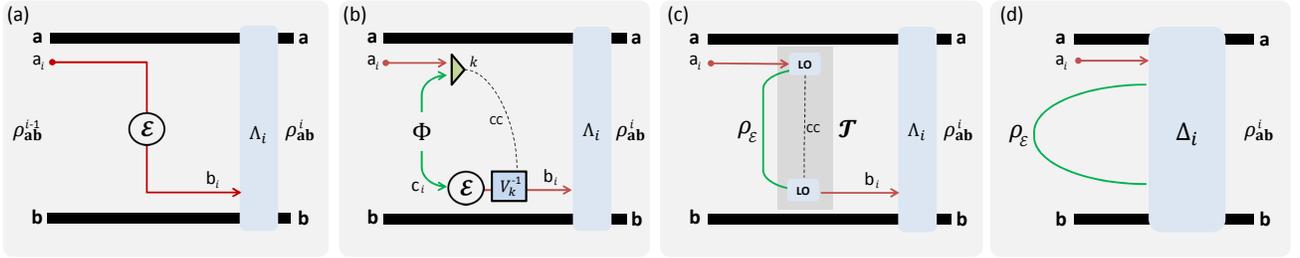} \vspace{-4cm}
\end{center}
\par
\vspace{-0.6cm}\caption{Teleportation stretching of an adaptive protocol.
(a)~Consider the $i$th transmission $a_{i}\rightarrow b_{i}$ through a
teleportation-covariant channel $\mathcal{E}$, followed by the QO $\Lambda
_{i}$, so that the register state $\rho_{\mathbf{ab}}^{i-1}:=\rho
_{\mathbf{a}a_{i}\mathbf{b}}$ is updated to $\rho_{\mathbf{ab}}^{i}$. (b) We
can replace the actual transmission with quantum teleportation. The input
system $a_{i}$ and part of a maximally-entangled state $\Phi$ are subject to a
Bell detection with outcome $k$. This process teleports the reduced state
$\rho_{a_{i}}$ of $a_{i}$ onto system $c_{i}$ up to a unitary operator $U_{k}%
$. Because the channel is teleportation-covariant, we have $\mathcal{E}%
(U_{k}\rho_{a_{i}}U_{k}{}^{\dagger})=V_{k}\mathcal{E}(\rho_{a_{i}}%
)V_{k}^{\dagger}$ so that $V_{k}$ can be undone at the output, and we retrieve
$\mathcal{E}(\rho_{a_{i}})$ on system $b_{i}$. This process also teleports all
the correlations that system $a_{i}$ may have with other systems in the
registers, i.e., it teleports part $a_{i}$ of the input state $\rho
_{\mathbf{a}a_{i}\mathbf{b}}$. (c)~Note that the propagation of $\Phi$ through
channel $\mathcal{E}$ defines its Choi matrix $\rho_{\mathcal{E}}$, and the
teleportation process over this state is just an LOCC, that becomes
trace-preserving after averaging over the Bell outcomes. In other words, we
may write $\mathcal{E}(\rho_{a_{i}})=\mathcal{T}(\rho_{a_{i}}\otimes
\rho_{\mathcal{E}})$ for a teleportation LOCC\ $\mathcal{T}$. This is a
particular case of Choi-stretchable channel as generally defined in
Ref.~\cite{PLOBapp}. Including the registers, we may write $\mathcal{I}%
_{\mathbf{a}}\otimes\mathcal{E}\otimes\mathcal{I}_{\mathbf{b}}(\rho
_{\mathbf{a}a_{i}\mathbf{b}})=\mathcal{I}_{\mathbf{a}}\otimes\mathcal{T}%
\otimes\mathcal{I}_{\mathbf{b}}(\rho_{\mathbf{a}a_{i}\mathbf{b}}\otimes
\rho_{\mathcal{E}})$. \ (d)~We finally collapse $\mathcal{I}_{\mathbf{a}%
}\otimes\mathcal{T}\otimes\mathcal{I}_{\mathbf{b}}$ and $\Lambda_{i}$ into a
single QO $\Delta_{i}$ applied to $\rho_{\mathbf{a}a_{i}\mathbf{b}}\otimes
\rho_{\mathcal{E}}$, so that we can write the recursive formula of
Eq.~(\ref{smain}).}%
\label{FIG4}%
\end{figure}

Here we explicitly show how to \textquotedblleft stretch\textquotedblright\ an
adaptive protocol of parameter estimation into a block form. This is a simple
adaptation of the general argument that Ref.~\cite{PLOBapp} originally
provided for protocols of quantum/private communication. We first consider
discrete-variable channels and then we extend the results to
continuous-variable channels afterwards. The procedure is explained in
Fig.~\ref{FIG4} for the $i$th transmission through an arbitrary
teleportation-covariant channel $\mathcal{E}$. As we can see, the register
state of the two parties is updated by the recursive formula
\begin{equation}
\rho_{\mathbf{ab}}^{i}=\Delta_{i}(\rho_{\mathcal{E}}\otimes\rho_{\mathbf{ab}%
}^{i-1}), \label{smain}%
\end{equation}
for some quantum operation (QO) $\Delta_{i}$. Iterating this formula for $n$
transmissions, we accumulate $n$ Choi matrices $\rho_{\mathcal{E}}^{\otimes
n}$ while collapsing the QOs. In our estimation protocol, after $n$ probings
of the channel $\mathcal{E}_{\theta}$, the register state becomes%
\begin{equation}
\rho_{\mathbf{ab}}^{n}(\theta)=\Delta\left(  \rho_{\mathcal{E}_{\theta}%
}^{\otimes n}\otimes\rho_{\mathbf{ab}}^{0}\right)  , \label{smain2}%
\end{equation}
where $\Delta=\Lambda_{n}\circ\cdots\circ\Lambda_{1}$ does not depend on
$\theta$. 

In Eq.~(\ref{smain2}), we may include the initial register state
$\rho_{\mathbf{ab}}^{0}$ into $\Delta$ and write
\begin{equation}
\rho_{\mathbf{ab}}^{n}(\theta)=\bar{\Lambda}\left(  \rho_{\mathcal{E}_{\theta
}}^{\otimes n}\right)  , \label{eq22}%
\end{equation}
for a trace-preserving and $\theta$-independent QO $\bar{\Lambda}$
(trace-preserving is assured by averaging over all measurements involved in
the teleportation simulation and the original adaptive protocol).

Note that we may repeat the reasoning in Fig.~\ref{FIG4} for a programmable
channel $\mathcal{E}$, which can be represented as in Fig.~\ref{FIG4}(c) but
with an arbitrary trace-preserving QO $\mathcal{U}$ (in the place of the
teleportation LOCC)\ applied to some programme state $\sigma_{\mathcal{E}}$
(in the place of the Choi matrix $\rho_{\mathcal{E}}$). This leads to a
different form of Eq.~(\ref{eq22}), namely%
\begin{equation}
\rho_{\mathbf{ab}}^{n}(\theta)=\tilde{\Lambda}\left(  \sigma_{\mathcal{E}%
_{\theta}}^{\otimes n}\right)  , \label{eq44}%
\end{equation}
for some other trace-preserving and $\theta$-independent QO $\tilde{\Lambda}$.

\subsection*{Extension to bosonic channels}

For a bosonic teleportation-covariant channel, we need to consider an
asymptotic simulation. In other words, we start from the imperfect simulation
$\mathcal{E}^{\mu}(\rho)=\mathcal{T}^{\mu}(\rho\otimes\rho_{\mathcal{E}}^{\mu
})$ where the teleportation LOCC $\mathcal{T}^{\mu}$ is built considering a
finite-energy POVM $\mathcal{B}^{\mu}$ (such that the ideal Bell detection is
defined as the limit $\mathcal{B}:=\lim_{\mu}\mathcal{B}^{\mu}$) and
$\rho_{\mathcal{E}}^{\mu}:=\mathcal{I}\otimes\mathcal{E}(\Phi^{\mu})$ defines
the bosonic Choi matrix as $\rho_{\mathcal{E}}:=\lim_{\mu}\rho_{\mathcal{E}%
}^{\mu}$. Because of the Braunstein-Kimble protocol~\cite{Samteleapp,DAriano},
for any bipartite state $\rho$, we have the point-wise limit
\begin{equation}
\Vert\mathcal{I}\otimes\mathcal{E}(\rho)-\mathcal{I}\otimes\mathcal{E}^{\mu
}(\rho)\Vert_{1}\overset{\mu}{\rightarrow}0~. \label{pointwise}%
\end{equation}
This limit can equivalently be expressed in terms of bounded diamond norm. In
fact, let us consider the (compact) set of energy-constrained bipartite states
$\mathcal{D}_{N}:=\{\rho~|~\mathrm{Tr}(\hat{N}\rho)\leq N\}$, where $\hat{N}$
is the total number operator. Then, for two bosonic channels, $\mathcal{E}%
_{1}$ and $\mathcal{E}_{2}$, one may define the bounded diamond
norm~\cite{PLOBapp}%
\begin{equation}
\left\Vert \mathcal{E}_{1}-\mathcal{E}_{2}\right\Vert _{\diamond N}%
:=\sup_{\rho\in\mathcal{D}_{N}}\Vert\mathcal{I}\otimes\mathcal{E}_{1}%
(\rho)-\mathcal{I}\otimes\mathcal{E}_{2}(\rho)\Vert_{1}~, \label{defBBBB}%
\end{equation}
which provides the standard (unbounded) diamond norm~\cite{Paulsenapp} in the
limit of large $N$, i.e.,
\begin{equation}
\left\Vert \mathcal{E}_{1}-\mathcal{E}_{2}\right\Vert _{\diamond}%
:=\lim_{N\rightarrow\infty}\left\Vert \mathcal{E}_{1}-\mathcal{E}%
_{2}\right\Vert _{\diamond N}~.
\end{equation}
By exploiting the fact that $\mathcal{D}_{N}$ is a compact set, the pointwise
limit in Eq.~(\ref{pointwise}) implies the uniform limit
\begin{equation}
\left\Vert \mathcal{E}-\mathcal{E}^{\mu}\right\Vert _{\diamond N}\overset{\mu
}{\rightarrow}0\text{~~\textrm{for any }}N. \label{defBBNN}%
\end{equation}
Therefore, for any $N<\infty$ and $\varepsilon>0$, there is a sufficiently
large $\mu$ such that $\Vert\mathcal{E}-\mathcal{E}^{\mu}\Vert_{\diamond
N}\leq\varepsilon$. For the estimation protocol this happens for any $\theta$,
so that we may write%
\begin{equation}
\Vert\mathcal{E}_{\theta}-\mathcal{E}_{\theta}^{\mu}\Vert_{\diamond N}%
\leq\varepsilon~. \label{primao}%
\end{equation}

The latter bound can be extended to the output of the adaptive protocol after
$n$ channel uses. Consider the original output state
\begin{equation}
\rho_{\mathbf{ab}}^{n}(\theta):=\Lambda_{n}\circ\mathcal{E}_{\theta}%
\circ\Lambda_{n-1}\cdots\circ\mathcal{E}_{\theta}(\rho_{\mathbf{ab}}^{0}),
\end{equation}
and its simulation
\begin{equation}
\rho_{\mathbf{ab}}^{n,\mu}(\theta):=\Lambda_{n}\circ\mathcal{E}_{\theta}^{\mu
}\circ\Lambda_{n-1}\cdots\circ\mathcal{E}_{\theta}^{\mu}(\rho_{\mathbf{ab}%
}^{0}),
\end{equation}
which is found by replacing $\mathcal{E}_{\theta}$ with $\mathcal{E}_{\theta
}^{\mu}$. Here it is understood that $\mathcal{E}_{\theta}$ and $\mathcal{E}%
_{\theta}^{\mu}$ are applied to system $a_{i}$ for the $i$-th transmission,
i.e., we have $\mathcal{E}_{\theta}=\mathcal{I}_{\mathbf{a}}\otimes
(\mathcal{E}_{\theta})_{a_{i}}\otimes\mathcal{I}_{\mathbf{b}}$. Assume that
the mean total number of photons in the states $\rho_{\mathbf{ab}}^{n}%
(\theta)$ and $\rho_{\mathbf{ab}}^{n,\mu}(\theta)$ is bounded by some large
but finite value $N(n)$ for any $\theta$ and $\mu$. Since these are physical
states, it is always possible to find such a common bound. In general, for $n$
uses, we have a sequence $\{N(0),\cdots,N(i),\cdots,N(n)\}$ of which $N(n)$
can always be chosen to be the greatest value.

Then, we may show that%
\begin{equation}
\Vert\rho_{\mathbf{ab}}^{n}(\theta)-\rho_{\mathbf{ab}}^{n,\mu}(\theta
)\Vert_{1}\leq n\left\Vert \mathcal{E}_{\theta}-\mathcal{E}_{\theta}^{\mu
}\right\Vert _{\diamond N(n)}~. \label{diamond2}%
\end{equation}
In fact, for $n=2$, we may write%
\begin{align}
\Vert\rho_{\mathbf{ab}}^{2}(\theta)-\rho_{\mathbf{ab}}^{2,\mu}(\theta
)\Vert_{1}  &  =\Vert\Lambda_{2}\circ\mathcal{E}_{\theta}\circ\Lambda_{1}%
\circ\mathcal{E}_{\theta}(\rho_{\mathbf{ab}}^{0})-\Lambda_{2}\circ
\mathcal{E}_{\theta}^{\mu}\circ\Lambda_{1}\circ\mathcal{E}_{\theta}^{\mu}%
(\rho_{\mathbf{ab}}^{0})\Vert_{1}\nonumber\\
&  \overset{(1)}{\leq}\Vert\mathcal{E}_{\theta}\circ\Lambda_{1}\circ
\mathcal{E}_{\theta}(\rho_{\mathbf{ab}}^{0})-\mathcal{E}_{\theta}^{\mu}%
\circ\Lambda_{1}\circ\mathcal{E}_{\theta}^{\mu}(\rho_{\mathbf{ab}}^{0}%
)\Vert_{1}\nonumber\\
&  \overset{(2)}{\leq}\Vert\mathcal{E}_{\theta}\circ\Lambda_{1}\circ
\mathcal{E}_{\theta}(\rho_{\mathbf{ab}}^{0})-\mathcal{E}_{\theta}\circ
\Lambda_{1}\circ\mathcal{E}_{\theta}^{\mu}(\rho_{\mathbf{ab}}^{0})\Vert
_{1}+\Vert\mathcal{E}_{\theta}\circ\Lambda_{1}\circ\mathcal{E}_{\theta}^{\mu
}(\rho_{\mathbf{ab}}^{0})-\mathcal{E}_{\theta}^{\mu}\circ\Lambda_{1}%
\circ\mathcal{E}_{\theta}^{\mu}(\rho_{\mathbf{ab}}^{0})\Vert_{1}\nonumber\\
&  \overset{(3)}{\leq}\Vert\mathcal{E}_{\theta}(\rho_{\mathbf{ab}}%
^{0})-\mathcal{E}_{\theta}^{\mu}(\rho_{\mathbf{ab}}^{0})\Vert_{1}%
+\Vert\mathcal{E}_{\theta}[\Lambda_{1}\circ\mathcal{E}_{\theta}^{\mu}%
(\rho_{\mathbf{ab}}^{0})]-\mathcal{E}_{\theta}^{\mu}[\Lambda_{1}%
\circ\mathcal{E}_{\theta}^{\mu}(\rho_{\mathbf{ab}}^{0})]\Vert_{1}\nonumber\\
&  \overset{(4)}{\leq}2\left\Vert \mathcal{E}_{\theta}-\mathcal{E}_{\theta
}^{\mu}\right\Vert _{\diamond N(n)}~, \label{casen2}%
\end{align}
where: (1)~we use the monotonicity under completely-positive trace-preserving
(CPTP) maps (note that the QO $\Lambda_{2}$ can always be made
trace-preserving by adding ancillas and delaying quantum measurements at the
end of the protocol); (2)~we use the triangle inequality; (3)~we use
monotonicity with respect to $\mathcal{E}_{\theta}\circ\Lambda_{1}$; and
(4)~we upperbound the trace distance via the bounded diamond norm. Extension
of Eq.~(\ref{casen2}) to arbitrary $n$ is just a matter of technicalities.

From Eq.~(\ref{primao}) we have that, for any finite $N(n)$ and $\varepsilon
>0$, there is a sufficiently large $\mu$ such that
\begin{equation}
\left\Vert \mathcal{E}_{\theta}-\mathcal{E}_{\theta}^{\mu}\right\Vert
_{\diamond N(n)}\leq\varepsilon~. \label{eqol}%
\end{equation}
Combining the latter with Eq.~(\ref{diamond2}) leads to%
\begin{equation}
\Vert\rho_{\mathbf{ab}}^{n}(\theta)-\rho_{\mathbf{ab}}^{n,\mu}(\theta
)\Vert_{1}\leq n\varepsilon~. \label{fromg}%
\end{equation}
By using a finite-energy simulation\ $\mathcal{T}^{\mu}$, we may may weaken
Eq.~(\ref{eq22}) into
\begin{equation}
\rho_{\mathbf{ab}}^{n,\mu}(\theta)=\bar{\Lambda}_{\mu}\left(  \rho
_{\mathcal{E}_{\theta}}^{\mu\otimes n}\right)  , \label{finitemu}%
\end{equation}
where the $\theta$-independent QO $\bar{\Lambda}_{\mu}$ is determined by the
original QOs of the protocol plus the teleportation LOCCs $\mathcal{T}^{\mu}$
($\bar{\Lambda}_{\mu}$ is trace-preserving by averaging over all
measurements). Thus, combining Eqs.~(\ref{fromg}) and (\ref{finitemu}), we
find that
\begin{equation}
\Vert\rho_{\mathbf{ab}}^{n}(\theta)-\bar{\Lambda}_{\mu}\left(  \rho
_{\mathcal{E}_{\theta}}^{\mu\otimes n}\right)  \Vert_{1}\leq n\varepsilon.
\end{equation}
or, equivalently, $\Vert\rho_{\mathbf{ab}}^{n}(\theta)-\bar{\Lambda}_{\mu
}\left(  \rho_{\mathcal{E}_{\theta}}^{\mu\otimes n}\right)  \Vert_{1}%
\overset{\mu}{\rightarrow}0$. Therefore, given an adaptive protocol with
arbitrary register energy $N(n)$, we may write its $n$-use\ output state as
the (trace-norm) limit%
\begin{equation}
\rho_{\mathbf{ab}}^{n}(\theta)=\lim_{\mu}\bar{\Lambda}_{\mu}\left(
\rho_{\mathcal{E}_{\theta}}^{\mu\otimes n}\right)  .
\end{equation}

\section{No-go: teleportation-covariance implies SQL (proof of Theorem~2)}

First consider discrete-variable\ teleportation-covariant
channels\ $\mathcal{E}_{\theta}$. Let us adopt the following notation
\begin{equation}
B(n,\theta):=\frac{8(1-F_{\theta}^{n})}{d\theta^{2}},~~F_{\theta}%
:=F(\rho_{\mathcal{E}_{\theta}},\rho_{\mathcal{E}_{\theta+d\theta}}).
\end{equation}
We first show that $B(n,\theta)$ is an upper bound for $\bar{I}_{\theta}^{n}$.
Given any adaptive protocol $\mathcal{P}$, we may write $\rho_{\mathbf{ab}%
}^{n}(\theta)=\bar{\Lambda}\left(  \rho_{\mathcal{E}_{\theta}}^{\otimes
n}\right)  $ with a $\theta$-independent QO $\bar{\Lambda}$. In particular,
this means that we may also write
\begin{equation}
\rho_{\mathbf{ab}}^{n}(\theta+d\theta)=\bar{\Lambda}\left(  \rho
_{\mathcal{E}_{\theta+d\theta}}^{\otimes n}\right)  .
\end{equation}
In order to bound the quantum Fisher information (QFI)
\begin{equation}
I_{\theta}^{n}(\mathcal{P})=\frac{8\left\{  1-F[\rho_{\mathbf{ab}}^{n}%
(\theta),\rho_{\mathbf{ab}}^{n}(\theta+d\theta)]\right\}  }{d\theta^{2}},
\label{1bisEQ}%
\end{equation}
we exploit basic properties of the quantum fidelity. In fact, we derive
\begin{equation}
F[\rho_{\mathbf{ab}}^{n}(\theta),\rho_{\mathbf{ab}}^{n}(\theta+d\theta
)]\overset{(1)}{\geq}F(\rho_{\mathcal{E}_{\theta}}^{\otimes n},\rho
_{\mathcal{E}_{\theta+d\theta}}^{\otimes n})\overset{(2)}{=}F(\rho
_{\mathcal{E}_{\theta}},\rho_{\mathcal{E}_{\theta+d\theta}})^{n}:=F_{\theta
}^{n}, \label{fidmm}%
\end{equation}
where we use: (1)~the monotonicity of the fidelity under CPTP maps, as is
$\bar{\Lambda}$; and (2)~its multiplicativity over tensor-product states.
Therefore, by using Eq.~(\ref{fidmm}) in Eq.~(\ref{1bisEQ}), we derive
$I_{\theta}^{n}(\mathcal{P})\leq B(n,\theta)$ for any protocol $\mathcal{P}$.
The latter bound is also valid for the supremum over all protocols, therefore
proving $\bar{I}_{\theta}^{n}\leq B(n,\theta)$.

The next step is to show that the bound $B(n,\theta)$ is additive. For $n=1$
and $d\theta\rightarrow0$, we may write $F_{\theta}=1-B(1,\theta)d\theta
^{2}/8$ which implies $F_{\theta}^{n}=1-nB(1,\theta)d\theta^{2}/8$ up to
$O(d\theta^{4})$. The latter expansion leads to $B(n,\theta)=nB(1,\theta)$, so
that we may directly write
\begin{equation}
\bar{I}_{\theta}^{n}\leq nB(1,\theta)=nB(\theta),~~B(\theta):=\frac
{8(1-F_{\theta})}{d\theta^{2}}. \label{kkkl}%
\end{equation}
Consider now a non-adaptive protocol $\mathcal{\tilde{P}}$ where Alice
prepares $n$ maximally-entangled (Bell) states $\Phi^{\otimes n}$ and partly
propagates them through the box, so that the output is $\rho_{\mathbf{ab}}%
^{n}(\theta)=\rho_{\mathcal{E}_{\theta}}^{\otimes n}$. By replacing this state
in Eq.~(\ref{1bisEQ}), we get $I_{\theta}^{n}(\mathcal{\tilde{P}})=nB(\theta
)$, so that $\bar{I}_{\theta}^{n}\geq nB(\theta)$. Combining the latter with
Eq.~(\ref{kkkl}) leads to $\bar{I}_{\theta}^{n}=nB(\theta)$. Since
$\mathcal{\tilde{P}}$ uses independent probing states, the quantum Cramer Rao
bound (QCRB) $\mathrm{Var}(\theta)\geq\lbrack I_{\theta}^{n}(\mathcal{\tilde
{P}})]^{-1}=[nB(\theta)]^{-1}$ is asymptotically achievable (for large $n$) by
using local measurements and adaptive estimators~\cite{Gillapp}.

For a programmable channel $\mathcal{E}_{\theta}$ with programme state
$\sigma_{\mathcal{E}_{\theta}}$ we may write $\rho_{\mathbf{ab}}^{n}%
(\theta)=\tilde{\Lambda}\left(  \sigma_{\mathcal{E}_{\theta}}^{\otimes
n}\right)  $, which leads to the following alternative version of
Eq.~(\ref{fidmm})%
\begin{equation}
F[\rho_{\mathbf{ab}}^{n}(\theta),\rho_{\mathbf{ab}}^{n}(\theta+d\theta)\geq
F(\sigma_{\mathcal{E}_{\theta}},\sigma_{\mathcal{E}_{\theta+d\theta}})^{n}.
\end{equation}
It is easy to repeat some of the previous steps to prove the bound
\begin{equation}
\bar{I}_{\theta}^{n}\leq n\frac{8\left[  1-F(\sigma_{\mathcal{E}_{\theta}%
},\sigma_{\mathcal{E}_{\theta+d\theta}})\right]  }{d\theta^{2}}. \label{pool}%
\end{equation}
However, we do not know if this bound is achievable or not, i.e., we cannot
put an equality in Eq.~(\ref{pool}), because we do not know if the programme
state $\sigma_{\mathcal{E}_{\theta}}$ can be generated by the transmission of
an input state through the channel.

\subsection*{Extension to bosonic channels}

Let us consider continuous-variable teleportation-covariant channels. For any
adaptive protocol $\mathcal{P}$, we may write%
\begin{equation}
I_{\theta}^{n}(\mathcal{P}):=4\frac{d_{B}^{2}[\rho_{\mathbf{ab}}^{n}%
(\theta),\rho_{\mathbf{ab}}^{n}(\theta+d\theta)]}{d\theta^{2}},
\label{buresPP}%
\end{equation}
where $d_{B}$ is the Bures distance%
\begin{equation}
d_{B}(\rho_{1},\rho_{2})=\sqrt{2[1-F(\rho_{1},\rho_{2})]}.
\end{equation}
The Bures distance between the output states, $\rho_{\mathbf{ab}}^{n}(\theta)$
and $\rho_{\mathbf{ab}}^{n}(\theta+d\theta)$, can be related to the Bures
distance between the $\mu$-approximate output states, $\rho_{\mathbf{ab}%
}^{n,\mu}(\theta)$ and $\rho_{\mathbf{ab}}^{n,\mu}(\theta+d\theta)$. In fact,
by applying the triangle inequality and bounding $d_{B}$ with the trace
distance $D$, i.e.,
\begin{equation}
d_{B}^{2}(\rho_{1},\rho_{2})\leq D(\rho_{1},\rho_{2}):=\frac{1}{2}||\rho
_{1}-\rho_{2}||_{1},
\end{equation}
we get the following
\begin{align}
d_{B}[\rho_{\mathbf{ab}}^{n}(\theta),\rho_{\mathbf{ab}}^{n}(\theta+d\theta)]
&  \leq d_{B}[\rho_{\mathbf{ab}}^{n}(\theta),\rho_{\mathbf{ab}}^{n,\mu}%
(\theta)]+d_{B}[\rho_{\mathbf{ab}}^{n,\mu}(\theta),\rho_{\mathbf{ab}}^{n,\mu
}(\theta+d\theta)]+d_{B}[\rho_{\mathbf{ab}}^{n,\mu}(\theta+d\theta
),\rho_{\mathbf{ab}}^{n}(\theta+d\theta)]\nonumber\\
&  \leq\sqrt{D[\rho_{\mathbf{ab}}^{n}(\theta),\rho_{\mathbf{ab}}^{n,\mu
}(\theta)]}+d_{B}[\rho_{\mathbf{ab}}^{n,\mu}(\theta),\rho_{\mathbf{ab}}%
^{n,\mu}(\theta+d\theta)]+\sqrt{D[\rho_{\mathbf{ab}}^{n,\mu}(\theta
+d\theta),\rho_{\mathbf{ab}}^{n}(\theta+d\theta)]}\nonumber\\
&  \leq\sqrt{\frac{n}{2}\left\Vert \mathcal{E}_{\theta}-\mathcal{E}_{\theta
}^{\mu}\right\Vert _{\diamond N(n)}}+d_{B}[\rho_{\mathbf{ab}}^{n,\mu}%
(\theta),\rho_{\mathbf{ab}}^{n,\mu}(\theta+d\theta)]+\sqrt{\frac{n}%
{2}\left\Vert \mathcal{E}_{\theta+d\theta}-\mathcal{E}_{\theta+d\theta}^{\mu
}\right\Vert _{\diamond N(n)}}~,
\end{align}
where, in the last step, we have also used Eq.~(\ref{diamond2}) with $N(n)$
being the energy bound of protocol $\mathcal{P}$.

Using Eq.~(\ref{eqol}) we see that, for any energy-bounded protocol
$\mathcal{P}$, there is a sufficiently large $\mu$ such that
\begin{equation}
d_{B}[\rho_{\mathbf{ab}}^{n}(\theta),\rho_{\mathbf{ab}}^{n}(\theta
+d\theta)]\leq\sqrt{2n\varepsilon}+d_{B}[\rho_{\mathbf{ab}}^{n,\mu}%
(\theta),\rho_{\mathbf{ab}}^{n,\mu}(\theta+d\theta)].
\end{equation}
In other words, we may write the following limit%
\begin{equation}
d_{B}[\rho_{\mathbf{ab}}^{n}(\theta),\rho_{\mathbf{ab}}^{n}(\theta
+d\theta)]\leq\lim_{\mu\rightarrow\infty}d_{B}[\rho_{\mathbf{ab}}^{n,\mu
}(\theta),\rho_{\mathbf{ab}}^{n,\mu}(\theta+d\theta)],
\end{equation}
which leads to%
\begin{equation}
I_{\theta}^{n}(\mathcal{P})\leq\lim_{\mu\rightarrow\infty}I_{\theta}^{n,\mu
}(\mathcal{P}),~~I_{\theta}^{n,\mu}(\mathcal{P}):=4\frac{d_{B}^{2}%
[\rho_{\mathbf{ab}}^{n,\mu}(\theta),\rho_{\mathbf{ab}}^{n,\mu}(\theta
+d\theta)]}{d\theta^{2}}.
\end{equation}
Now note that, at any finite $\mu$, we can use Eq.~(\ref{finitemu}) and%
\begin{equation}
\rho_{\mathbf{ab}}^{n,\mu}(\theta+d\theta)=\bar{\Lambda}_{\mu}\left(
\rho_{\mathcal{E}_{\theta+d\theta}}^{\mu\otimes n}\right)  .
\end{equation}
It is then easy to see that the derivation in Eq.~(\ref{fidmm}) can be
modified into%
\begin{equation}
F[\rho_{\mathbf{ab}}^{n,\mu}(\theta),\rho_{\mathbf{ab}}^{n,\mu}(\theta
+d\theta)]\geq(F_{\theta}^{\mu})^{n}:=F(\rho_{\mathcal{E}_{\theta}}^{\mu}%
,\rho_{\mathcal{E}_{\theta+d\theta}}^{\mu})^{n}.
\end{equation}
Therefore, for any energy-bounded protocol $\mathcal{P}$, we may write the
following bound for the $\mu$-dependent QFI%
\begin{equation}
I_{\theta}^{n,\mu}(\mathcal{P})\leq B(n,\theta,\mu):=\frac{8[1-(F_{\theta
}^{\mu})^{n}]}{d\theta^{2}}.
\end{equation}
As before, it is immediate to prove the additivity, so that we derive%
\begin{equation}
I_{\theta}^{n,\mu}(\mathcal{P})\leq nB(\theta,\mu),~~B(\theta,\mu
):=\frac{8\left[  1-F(\rho_{\mathcal{E}_{\theta}}^{\mu},\rho_{\mathcal{E}%
_{\theta+d\theta}}^{\mu})\right]  }{d\theta^{2}}~. \label{seegg}%
\end{equation}

By taking the limit for large $\mu$ and optimizing over all $\mathcal{P}$, we
therefore get
\begin{equation}
\bar{I}_{\theta}^{n}:=\sup_{\mathcal{P}}I_{\theta}^{n}(\mathcal{P})\leq
\lim_{\mu\rightarrow\infty}nB(\theta,\mu)=n\frac{8\left[  1-\lim_{\mu}%
F(\rho_{\mathcal{E}_{\theta}}^{\mu},\rho_{\mathcal{E}_{\theta+d\theta}}^{\mu
})\right]  }{d\theta^{2}}~. \label{bbnn}%
\end{equation}
Note that, because we consider a supremum in the definition of $\bar
{I}_{\theta}^{n}$, we may also include the limit of energy-unbounded
protocols. As a matter of fact, such asymptotic protocols are those saturating
the upper bound. In fact, consider a non-adaptive protocol $\mathcal{\tilde
{P}}_{\mu}$, where Alice transmits part of two-mode squeezed vacuum (TMSV)
states $\Phi^{\mu\otimes n}$, so that the $n$-use output state is
$\rho_{\mathbf{ab}}^{n}(\theta)=\rho_{\mathcal{E}_{\theta}}^{\mu\otimes n}$.
By replacing the latter in Eq.~(\ref{buresPP}), we derive $I_{\theta}%
^{n}(\mathcal{\tilde{P}}_{\mu})=nB(\theta,\mu)$. By taking the limit for large
$\mu$, we define an asymptotic protocol $\mathcal{\tilde{P}}:=\lim_{\mu
}\mathcal{\tilde{P}}_{\mu}$ with asymptotic performance
\begin{equation}
I_{\theta}^{n}(\mathcal{\tilde{P}}):=\lim_{\mu\rightarrow\infty}I_{\theta}%
^{n}(\mathcal{\tilde{P}}_{\mu})=\lim_{\mu\rightarrow\infty}nB(\theta,\mu),
\end{equation}
which achieves the upper bound in Eq.~(\ref{bbnn}). Since $\mathcal{\tilde{P}%
}_{\mu}$ (and its limit $\mathcal{\tilde{P}}$)\ uses independent probing
states, the corresponding quantum Cramer Rao bound (QCRB) is achievable for
large $n$.

\section{Limits of multiparameter adaptive noise estimation}

\subsection*{Preliminaries}

Consider a quantum state $\rho$ which is function of a multiparameter
$\Theta=(\theta^{1},\theta^{2},\dots,\theta^{m})$. Let $\mathbb{I}_{\Theta}$
be the corresponding QFI matrix. Its elements are expressed in terms of the
symmetric logarithmic derivative $L_{\mu}$ as follows~\cite{Parisapp}%
\begin{equation}
\mathbb{I}_{\Theta}^{\mu\nu}=\mathrm{Tr}\left(  \rho\frac{L_{\mu}L_{\nu
}+L_{\nu}L_{\mu}}{2}\right)  ,~~L_{\mu}:=\sum_{jk\,|\,D_{j}+D_{k}>0}\frac
{2}{D_{j}+D_{k}}\langle e_{j}|\frac{\partial\rho}{\partial\theta^{\mu}}%
|e_{k}\rangle|e_{j}\rangle\langle e_{k}|, \label{SLDdef}%
\end{equation}
where $\{|e_{k}\rangle\}$ are the eigenvectors of $\rho$ and $\{D_{k}\}$ its
eigenvalues. After $n$ rounds,\ the QCRB takes the form~\cite{Parisapp}%
\begin{equation}
\mathrm{Cov}(\hat{\Theta})\geq\frac{\mathbb{I}_{\Theta}^{-1}}{n},
\end{equation}
where $\mathrm{Cov}(\hat{\Theta})_{\mu\nu}:=\langle\hat{\theta}^{\mu}%
\hat{\theta}^{\nu}\rangle-\langle\hat{\theta}^{\mu}\rangle\langle\hat{\theta
}^{\nu}\rangle$ is the covariance matrix for the optimal multiparameter
estimator $\hat{\Theta}$. In the general scenario of joint multiparameter
estimation, the previous QCRB is not known to be achievable.

Consider now a curve in the parameter space $\theta^{\mu}(\tau)$. The quantum
estimation of parameter $\tau$ is bounded by a corresponding QFI $I_{\tau
}=\mathrm{Tr}(\rho L_{\tau}^{2})$ where $L_{\tau}$ is defined as in
Eq.~(\ref{SLDdef}) but with the replacement $\partial\theta^{\mu}%
\rightarrow\partial\tau$. From the relation
\begin{equation}
\frac{\partial}{\partial\tau}=\sum_{\mu=1}^{m}\frac{\partial\theta^{\mu}%
}{\partial\tau}\frac{\partial}{\partial\theta^{\mu}}=\sum_{\mu=1}^{m}%
\dot{\theta}^{\mu}\,\frac{\partial}{\partial\theta^{\mu}}~,
\end{equation}
we obtain its QFI in terms of the QFI matrix
\begin{equation}
I_{\tau}=\sum_{\mu\nu}\mathbb{I}_{\Theta}^{\mu\nu}\dot{\theta}^{\mu}%
\dot{\theta}^{\nu}~. \label{FF}%
\end{equation}

\subsection*{QFI matrix for adaptive protocols}

Consider now a teleportation-covariant channel $\mathcal{E}_{\Theta}$
depending on the multiparameter $\Theta=\{\theta^{\mu}\}$. This channel can
also be expressed in terms of the single parameter $\tau$ which defines the
curve $\theta^{\mu}(\tau)$. Given $n$ uses of an arbitrary adaptive protocol
$\mathcal{P}$, we consider the QFI matrix $\mathbb{I}_{\Theta}(\rho
_{\mathbf{ab}}^{n})$ associated with the estimation of $\Theta$ in the output
state $\rho_{\mathbf{ab}}^{n}$. We also consider the QFI $I_{\tau}%
(\rho_{\mathbf{ab}}^{n})$ associated with the estimation of the parameter
$\tau$. Note that we may write
\begin{equation}
I_{\tau}(\rho_{\mathbf{ab}}^{n})=\sum_{\mu\nu}\mathbb{I}_{\Theta}%
(\rho_{\mathbf{ab}}^{n})\dot{\theta}^{\mu}\dot{\theta}^{\nu}~. \label{cosmo1}%
\end{equation}
Because the channel is teleportation-covariant, we may also write
\begin{equation}
I_{\tau}(\rho_{\mathbf{ab}}^{n})\leq nI_{\tau}(\rho_{\mathcal{E}})~,
\label{Res}%
\end{equation}
where $I_{\tau}(\rho_{\mathcal{E}})$ is the QFI associated with the estimation
of parameter $\tau$ encoded in the channel's Choi matrix. Similarly, we may
write
\begin{equation}
I_{\tau}(\rho_{\mathcal{E}})=\sum_{\mu\nu}\mathbb{I}_{\Theta}^{\mu\nu}%
(\rho_{\mathcal{E}})\dot{\theta}^{\mu}\dot{\theta}^{\nu}~, \label{cosmo2}%
\end{equation}
where $\mathbb{I}_{\Theta}(\rho_{\mathcal{E}})$ is the QFI matrix associated
with the estimation of $\Theta$ in the Choi matrix.

From Eqs.~(\ref{cosmo1}), ~(\ref{Res}) and~(\ref{cosmo2}), we obtain
\begin{equation}
\sum_{\mu\nu}\mathbb{I}_{\Theta}^{\mu\nu}(\rho_{\mathbf{ab}}^{n})\dot{\theta
}^{\mu}\dot{\theta}^{\nu}\leq n\sum_{\mu\nu}\mathbb{I}_{\Theta}^{\mu\nu}%
(\rho_{\mathcal{E}})\dot{\theta}^{\mu}\dot{\theta}^{\nu}~.
\end{equation}
Since this is true for all $\dot{\theta}^{\mu}$, we finally obtain the QFI
matrix inequality
\begin{equation}
\mathbb{I}_{\Theta}(\rho_{\mathbf{ab}}^{n})\leq n\mathbb{I}_{\Theta}%
(\rho_{\mathcal{E}})~,
\end{equation}
which is valid for any adaptive protocol $\mathcal{P}$. Clearly, we still have
the SQL scaling.

\section{Computing the adaptive QFI for Pauli, erasure and Gaussian
channels\label{app2}}

First consider a qudit generalized Pauli channel $\mathcal{E}_{\boldsymbol{p}%
}$ with probability distribution $\boldsymbol{p}:=\{p_{k}\}$. This is
described by
\begin{equation}
\rho\rightarrow\mathcal{E}_{\boldsymbol{p}}(\rho)=\sum_{k=0}^{d^{2}-1}%
p_{k}P_{k}\rho P_{k}^{\dagger},
\end{equation}
where $P_{k}$ are a collection of $d^{2}$ generalized Pauli
operators~\cite{PLOBapp}. Its Choi matrix is given by%
\begin{equation}
\rho_{\mathcal{E}_{\boldsymbol{p}}}=\sum_{k=0}^{d^{2}-1}p_{k}\beta_{k},
\end{equation}
where $\beta_{k}=(I\otimes P_{k})\Phi(I\otimes P_{k})$ are the projectors over
the elements of a generalized Bell basis, with $\Phi=\left\vert \Phi
\right\rangle \left\langle \Phi\right\vert $ and
\begin{equation}
\left\vert \Phi\right\rangle :=d^{-1/2}\sum_{j=0}^{d-1}\left\vert
jj\right\rangle .
\end{equation}

Given two qudit Pauli channels, $\mathcal{E}_{0}:=\mathcal{E}_{\boldsymbol{p}%
^{0}}$ and $\mathcal{E}_{1}:=\mathcal{E}_{\boldsymbol{p}^{1}}$, with
probability distributions $\boldsymbol{p}^{0}=\{p_{k}^{0}\}$ and
$\boldsymbol{p}^{1}=\{p_{k}^{1}\}$, the Bures' fidelity between their Choi
matrices reads
\begin{equation}
F(\rho_{\mathcal{E}_{0}},\rho_{\mathcal{E}_{1}})=F(\boldsymbol{p}%
^{0},\boldsymbol{p}^{1}):=\sum_{k=0}^{d^{2}-1}\sqrt{p_{k}^{0}p_{k}^{1}}.
\end{equation}
For $\boldsymbol{p}^{0}=\boldsymbol{p}$ and $\boldsymbol{p}^{1}=\boldsymbol{p}%
+\boldsymbol{\delta p}$ (with $\sum_{k=0}^{d^{2}-1}\delta p_{k}=0$), we
derive
\begin{equation}
F(\rho_{\mathcal{E}_{\boldsymbol{p}}},\rho_{\mathcal{E}_{\boldsymbol{p}%
+\boldsymbol{\delta p}}})=F(\boldsymbol{p},\boldsymbol{p}+\boldsymbol{\delta
p})\simeq1-\frac{1}{8}\sum_{k=0}^{d^{2}-1}\frac{\delta p_{k}^{2}}{p_{k}}.
\label{toREP2}%
\end{equation}

As an example, consider a qubit depolarizing channel~\cite{Nielsenapp}, so
that we have $\boldsymbol{p}=\{1-p,p/3,p/3,p/3\}$ and $\boldsymbol{\delta
p}=\{-\delta p,\delta p/3,\delta p/3,\delta p/3\}$. By replacing in
Eq.~(\ref{toREP2}), we get
\begin{equation}
F(\boldsymbol{p},\boldsymbol{p}+\boldsymbol{\delta p})\simeq1-\frac{\delta
p^{2}}{8}\frac{1}{p(1-p)}.
\end{equation}
The latter equation leads to the following QFI
\begin{equation}
B(p):=\frac{8[1-F(\boldsymbol{p},\boldsymbol{p}+\boldsymbol{\delta p}%
)]}{\delta p^{2}}=\frac{1}{p(1-p)}.
\end{equation}
The same expression holds for the dephasing channel~\cite{Nielsenapp}, for
which $\boldsymbol{p}=\{1-p,0,0,p\}$ and $\boldsymbol{\delta p}=\{-\delta
p,0,0,\delta p\}$.

Let us now consider the qudit erasure channel
\begin{equation}
\rho\rightarrow\mathcal{E}_{\pi}(\rho)=(1-\pi)\rho+\pi\left\vert
e\right\rangle \left\langle e\right\vert ,
\end{equation}
where $\left\vert e\right\rangle $ is an erasure state, picked with
probability $\pi$. The corresponding Choi matrix is
\begin{equation}
\rho_{\mathcal{E}_{\pi}}=(1-\pi)\Phi+\pi\frac{I}{d}\otimes\left\vert
e\right\rangle \left\langle e\right\vert .
\end{equation}
The Bures' fidelity between two erasure channels, with different probabilities
$\pi$ and $\pi^{\prime}$, reads
\begin{equation}
F(\rho_{\mathcal{E}_{\pi}},\rho_{\mathcal{E}_{\pi^{\prime}}})=\sqrt
{(1-\pi)(1-\pi^{\prime})}+\sqrt{\pi\pi^{\prime}}.
\end{equation}
Setting $\pi=p$ and $\pi^{\prime}=p+\delta p$ we can easily compute the QFI
for the estimation of the erasure probability $p$, which is given by the same
expression found before, i.e.,%
\begin{equation}
B(p)=\left[  p(1-p)\right]  ^{-1}~.
\end{equation}

Bosonic Gaussian channels are uniquely determined by their action on the first
and second moments of the quadrature operators. In particular, the
thermal-loss channel and the amplifier channel transform the covariance
matrix~\cite{WeeRMPapp} $V$ of an input state as follows
\begin{equation}
V\rightarrow\eta V+|1-\eta|(\bar{n}_{T}+1/2),
\end{equation}
where $\eta$ is a real gain parameter and $\bar{n}_{T}$ is the mean number of
thermal photons in the environment. The thermal-loss channel is obtained for
$\eta\in\lbrack0,1)$, while the noisy amplifier for $\eta>1$. In both cases
our goal is to estimate the value of the positive noise parameter
$\theta\equiv\bar{n}_{T}>0$ for any fixed gain $\eta$. It is easy to check
that the state $\rho_{\mathcal{E}_{\theta}}^{\mu}:=\rho^{\mu}(\eta,\bar{n}%
_{T})$ is a Gaussian state with zero mean and covariance matrix
\begin{equation}
V_{\eta,\bar{n}_{T},\mu}=\left(
\begin{array}
[c]{cccc}%
\mu & 0 & \sqrt{\eta(\mu^{2}-1/4)} & 0\\
0 & \mu & 0 & -\sqrt{\eta(\mu^{2}-1/4)}\\
\sqrt{\eta(\mu^{2}-1/4)} & 0 & \eta\mu+|1-\eta|(\bar{n}_{T}+1/2) & 0\\
0 & -\sqrt{\eta(\mu^{2}-1/4)} & 0 & \eta\mu+|1-\eta|(\bar{n}_{T}+1/2)
\end{array}
\right)  . \label{ea}%
\end{equation}
By using the formula for the fidelity of multimode Gaussian
states~\cite{Banchiapp}, it is immediate to compute the $\mu$-dependent QFI
for the estimation of $\bar{n}_{T}$. For any protocol $\mathcal{P}$, we have
[from Eq.~(\ref{seegg})]
\begin{equation}
I_{\bar{n}_{T}}^{n,\mu}(\mathcal{P})\leq nB(\bar{n}_{T},\mu),~~B(\bar{n}%
_{T},\mu)=\frac{8\{1-F[\rho^{\mu}(\eta,\bar{n}_{T}),\rho^{\mu}(\eta,\bar
{n}_{T}+d\bar{n}_{T})]\}}{d\bar{n}_{T}^{2}}.
\end{equation}
Explicitly, we compute%
\begin{equation}
B(\bar{n}_{T},\mu)=\frac{1}{\bar{n}_{T}(\bar{n}_{T}+1)}\,\frac{|1-\eta
|(2+4\bar{n}_{T})\mu+1-\eta}{|1-\eta|(2+4\bar{n}_{T})\mu+1+\eta}.
\end{equation}
Therefore, by taking the limit for large $\mu$ and optimizing over all
protocols, we derive%
\begin{equation}
\bar{I}_{\bar{n}_{T}}^{n}=n\lim_{\mu}B(\bar{n}_{T},\mu)=\frac{n}{\bar{n}%
_{T}(\bar{n}_{T}+1)}~. \label{QFIooo}%
\end{equation}

Now consider the additive-noise Gaussian channel, which transforms the input
covariance matrix as $V\rightarrow V+wI$. For any $w>0$, the covariance matrix
of the state $\rho_{\mathcal{E}_{\theta}}^{\mu}:=\rho^{\mu}(w)$ reads
\begin{equation}
V_{w,\mu}=\left(
\begin{array}
[c]{cccc}%
\mu & 0 & \sqrt{\mu^{2}-1/4} & 0\\
0 & \mu & 0 & -\sqrt{\mu^{2}-1/4}\\
\sqrt{\mu^{2}-1/4} & 0 & \mu+w & 0\\
0 & -\sqrt{\mu^{2}-1/4} & 0 & \mu+w
\end{array}
\right)  .
\end{equation}
After simple algebra we compute the $\mu$-dependent QFI. For any protocol, we
have%
\begin{equation}
I_{w}^{n,\mu}(\mathcal{P})\leq nB(w,\mu),~~B(w,\mu)=\frac{8\mu}{8w^{2}\mu+4w},
\end{equation}
which leads to the adaptive QFI%
\begin{equation}
\bar{I}_{w}^{n}=n\lim_{\mu}B(w,\mu)=nw^{-2}~.
\end{equation}

\section{Limits for adaptive quantum channel discrimination (proof of
Theorem~3)}

First consider teleportation-covariant channels in finite dimension
(discrete-variable channels). Let us use the decomposition $\rho_{\mathbf{ab}%
}^{n}(k)=\bar{\Lambda}\left(  \rho_{\mathcal{E}_{k}}^{\otimes n}\right)  $ in
the protocol-dependent error probability%
\begin{equation}
p(k^{\prime}\neq k|\mathcal{P})=\frac{1-D[\rho_{\mathbf{ab}}^{n}%
(0),\rho_{\mathbf{ab}}^{n}(1)]}{2}. \label{appPmin}%
\end{equation}
Then, we may write%
\begin{equation}
D[\rho_{\mathbf{ab}}^{n}(0),\rho_{\mathbf{ab}}^{n}(1)]\leq D(\rho
_{\mathcal{E}_{0}}^{\otimes n},\rho_{\mathcal{E}_{1}}^{\otimes n}),
\label{discrgg}%
\end{equation}
where we use the monotonicity of the trace distance under the CPTP map
$\bar{\Lambda}$. We do not simplify $D(\rho_{\mathcal{E}_{\theta}}^{\otimes
n},\rho_{\mathcal{E}_{\theta+d\theta}}^{\otimes n})\leq nD(\rho_{\mathcal{E}%
_{\theta}},\rho_{\mathcal{E}_{\theta+d\theta}})$ because the bound may become
too large. Replacing Eq.~(\ref{discrgg}) in Eq.~(\ref{appPmin}), we get
\begin{equation}
p(k^{\prime}\neq k|\mathcal{P})\geq H_{n}:=[1-D(\rho_{\mathcal{E}_{0}%
}^{\otimes n},\rho_{\mathcal{E}_{1}}^{\otimes n})]/2,
\end{equation}
for any protocol $\mathcal{P}$, which is automatically extended to the infimum
over all protocols, thus proving $p_{\mathrm{err}}\geq H_{n}$ (in particular,
the infimum is a minimum in the discrete-variable case). To show that the
bound $H_{n}$ is achievable, consider a non-adaptive protocol $\mathcal{\tilde
{P}}$, where Alice prepares $n$ maximally-entangled (Bell) states
$\Phi^{\otimes n}$ and partly propagates them through the box, so that the
output state is equal to $\rho_{\mathbf{ab}}^{n}(k)=\rho_{\mathcal{E}_{k}%
}^{\otimes n}$. By replacing this output state in Eq.~(\ref{appPmin}), we
obtain $p(k^{\prime}\neq k|\mathcal{\tilde{P}})=H_{n}$. Therefore, we may
write $p_{\mathrm{err}}=H_{n}$.

Let us note that, in general, for two programmable channels $\mathcal{E}_{0}$
and $\mathcal{E}_{1}$, with programme states $\sigma_{\mathcal{E}_{0}}$ and
$\sigma_{\mathcal{E}_{1}}$, we may also write the decomposition $\rho
_{\mathbf{ab}}^{n}(k)=\tilde{\Lambda}\left(  \sigma_{\mathcal{E}_{k}}^{\otimes
n}\right)  $ for some CPTP map $\tilde{\Lambda}$. By repeating the previous
derivation, we therefore get%
\begin{equation}
p(k^{\prime}\neq k|\mathcal{P})\geq\lbrack1-D(\sigma_{\mathcal{E}_{0}%
}^{\otimes n},\sigma_{\mathcal{E}_{1}}^{\otimes n})]/2,
\end{equation}
for any protocol $\mathcal{P}$. This lower bound also applies to the infimum
$p_{\mathrm{err}}$. However, in general, we do not know if $p_{\mathrm{err}}$
is achievable, because it is not automatically guaranteed that the programme
states $\sigma_{\mathcal{E}_{k}}$\ can be generated by the transmission of
some input state through the channels $\mathcal{E}_{k}$.

\subsection*{Teleportation-covariance and diamond norm}

An adaptive protocol for the symmetric discrimination of two equiprobable
channels $\mathcal{E}_{0}$ and $\mathcal{E}_{1}$ represents a more general
strategy with respect to the block strategy of: (i)~preparing an arbitrary
input state $\rho_{AB}$, where the input system $A$ is generally
entangled/correlated with an ancillary system $B$; (ii)~sending $A$ through
the unknown channel $\mathcal{E}_{k}^{\otimes n}$, and (iii)~finally making an
optimal POVM jointly on the output of $A$ and the ancillary $B$. For this
reason, the adaptive minimum error probability $p_{\mathrm{err}}$ lowerbounds
the error probability associated with the optimization over all such block
strategies. In other words, for arbitrary $n$ uses, we may write
\begin{equation}
p_{\mathrm{err}}\leq\frac{1-\frac{1}{2}||\mathcal{E}_{0}^{\otimes
n}-\mathcal{E}_{1}^{\otimes n}||_{\diamond}}{2},~~||\mathcal{E}_{0}^{\otimes
n}-\mathcal{E}_{1}^{\otimes n}||_{\diamond}:=\sup_{\rho_{AB}}||\mathcal{E}%
_{0}^{\otimes n}\otimes\mathcal{I}_{B}(\rho_{AB})-\mathcal{E}_{1}^{\otimes
n}\otimes\mathcal{I}_{B}(\rho_{AB})||_{1}~. \label{pminKK}%
\end{equation}
However, we have previously proven that a specific type of block protocol
$\mathcal{\tilde{P}}$, based on maximally-entangled states $\Phi^{\otimes n}$
at the input (and, therefore, Choi matrices $\rho_{\mathcal{E}_{k}}^{\otimes
n}$ at the output) is able to reach the ultimate bound $p_{\mathrm{err}}$. As
a result, Eq.~(\ref{pminKK}) must hold with an equality, i.e., we may write%
\begin{equation}
p_{\mathrm{err}}=\frac{1-\frac{1}{2}||\mathcal{E}_{0}^{\otimes n}%
-\mathcal{E}_{1}^{\otimes n}||_{\diamond}}{2},
\end{equation}
for the adaptive discrimination of any pair of teleportation-covariant
channels in finite dimension.

\subsection*{Extension to bosonic channels}

The proof can be extended to bosonic teleportation-covariant channels by using
the finite-energy decomposition
\begin{equation}
\rho_{\mathbf{ab}}^{n,\mu}(k)=\bar{\Lambda}_{\mu}\left(  \rho_{\mathcal{E}%
_{k}}^{\mu\otimes n}\right)  , \label{eq73}%
\end{equation}
which is obtained by stretching the protocol via a finite-energy teleportation
LOCC $\mathcal{T}^{\mu}$. We may repeat the same reasoning that leads to
Eq.~(\ref{diamond2}) and obtain%
\begin{equation}
\Vert\rho_{\mathbf{ab}}^{n}(k)-\rho_{\mathbf{ab}}^{n,\mu}(k)\Vert_{1}\leq
n\Vert\mathcal{E}_{k}-\mathcal{E}_{k}^{\mu}\Vert_{\Diamond N(n)}~,
\end{equation}
for any adaptive protocol with arbitrary energy bound $N(n)$ (as previously
defined, this is a bound on the mean total number of photons present in the
registers at step $n$ for both the original and simulated protocol). For any
finite $N(n)$ and $\varepsilon>0$, there is a sufficiently large value of
$\mu$, such that
\begin{equation}
\Sigma_{\mu}:=\Vert\mathcal{E}_{k}-\mathcal{E}_{k}^{\mu}\Vert_{\Diamond
N(n)}\leq\varepsilon~,
\end{equation}
as a consequence of the Braunstein-Kimble protocol~\cite{Samteleapp,DAriano},
as also discussed in Sec.~\ref{app1} for the case of a continuous parameter
$\theta$; in particular, see Eq.~(\ref{pointwise}) implying Eq.~(\ref{eqol}).

Using the triangle inequality, we may write the following bound for the trace
distance%
\begin{align}
D[\rho_{\mathbf{ab}}^{n}(0),\rho_{\mathbf{ab}}^{n}(1)]  &  \leq D[\rho
_{\mathbf{ab}}^{n}(0),\rho_{\mathbf{ab}}^{n,\mu}(0)]+D[\rho_{\mathbf{ab}%
}^{n,\mu}(0),\rho_{\mathbf{ab}}^{n,\mu}(1)]+D[\rho_{\mathbf{ab}}^{n,\mu
}(1),\rho_{\mathbf{ab}}^{n}(1)]\nonumber\\
&  \leq n\Sigma_{\mu}+D[\rho_{\mathbf{ab}}^{n,\mu}(0),\rho_{\mathbf{ab}%
}^{n,\mu}(1)]
\end{align}
As a consequence, for any energy-bounded protocol $\mathcal{P}$, we may write%
\begin{equation}
p(k^{\prime}\neq k|\mathcal{P})\geq\frac{1-n\Sigma_{\mu}-D[\rho_{\mathbf{ab}%
}^{n,\mu}(0),\rho_{\mathbf{ab}}^{n,\mu}(1)]}{2}\geq\frac{1-n\Sigma_{\mu
}-D[\rho_{\mathcal{E}_{0}}^{\mu\otimes n},\rho_{\mathcal{E}_{1}}^{\mu\otimes
n}]}{2},
\end{equation}
where the last inequality exploits Eq.~(\ref{eq73}) combined with the
monotonicity of the trace distance under the CPTP map $\bar{\Lambda}_{\mu}$.
In the limit of large $\mu$, $\Sigma_{\mu}$ goes to zero, so that we achieve
perfect simulation and we may write\
\begin{equation}
p(k^{\prime}\neq k|\mathcal{P})\geq\frac{1-\lim_{\mu}D[\rho_{\mathcal{E}_{0}%
}^{\mu\otimes n},\rho_{\mathcal{E}_{1}}^{\mu\otimes n}]}{2}. \label{PforBB}%
\end{equation}

Since the optimal value $p_{\mathrm{err}}$ is defined as an infimum, we may
extend the lower bound in Eq.~(\ref{PforBB}) to the asymptotic limit of
energy-unbounded protocols (i.e., to the limit of large $N(n)$). Thus, for any
$n$ we may write
\begin{equation}
p_{\mathrm{err}}\geq\frac{1-\lim_{\mu}D[\rho_{\mathcal{E}_{0}}^{\mu\otimes
n},\rho_{\mathcal{E}_{1}}^{\mu\otimes n}]}{2}. \label{eqlimm}%
\end{equation}
Indeed the achievability of the latter bound is asymptotic. We consider a
non-adaptive protocol $\mathcal{\tilde{P}}$, where Alice prepares $n$ TMSV
states $\Phi^{\mu\otimes n}$ and partly propagates them through the box, so
that the output state is equal to $\rho_{\mathbf{ab}}^{n}(k)=\rho
_{\mathcal{E}_{k}}^{\mu\otimes n}$. By performing an optimal POVM, we get%
\begin{equation}
p(k^{\prime}\neq k|\mathcal{\tilde{P}})=\frac{1-D(\rho_{\mathcal{E}_{0}}%
^{\mu\otimes n},\rho_{\mathcal{E}_{1}}^{\mu\otimes n})}{2},
\end{equation}
which coincides with the lower bound of Eq.~(\ref{eqlimm}) in the limit of
large $\mu$.

\section{Single-letter bounds for adaptive quantum channel discrimination}

In Eq.~(14) of the main text, we provide various single-letter bounds for the
adaptive error probability $p_{\mathrm{err}}$. The fidelity bounds come from
the Fuchs-van der Graaf relations~\cite{Fuchsapp} between the Bures fidelity
$F$ and the trace distance $D$. For any two states, $\rho$ and $\sigma$, one
has~\cite{Fuchsapp}
\begin{equation}
1-F(\rho,\sigma)\leq D(\rho,\sigma)\leq\sqrt{1-F(\rho,\sigma)^{2}}.
\end{equation}
The minimum average error probability for discriminating two equiprobable
states $\rho$ and $\sigma$ is the Helstrom bound~\cite{Hesltromapp}%
\ $p(\rho\neq\sigma)=[1-D(\rho,\sigma)]/2$. Therefore, the previous relations
lead to
\begin{equation}
\frac{1-\sqrt{1-F(\rho,\sigma)^{2}}}{2}\leq p(\rho\neq\sigma)\leq\frac
{F(\rho,\sigma)}{2}. \label{FvDG}%
\end{equation}
Using the multiplicativity of the fidelity over tensor products, we may extend
Eq.~(\ref{FvDG}) to $n$-copy discrimination%
\begin{equation}
\frac{1-\sqrt{1-F(\rho,\sigma)^{2n}}}{2}\leq p(\rho^{\otimes n}\neq
\sigma^{\otimes n})\leq\frac{F(\rho,\sigma)^{n}}{2}. \label{pbbounds}%
\end{equation}

In our work we show that, for a pair of equiprobable teleportation-covariant
channels, $\mathcal{E}_{0}$ and $\mathcal{E}_{1}$, the adaptive error
probability is equal to the mimimum average error probability associated with
the discrimination of their Choi matrices
\begin{equation}
p_{\mathrm{err}}=p(\rho_{\mathcal{E}_{0}}^{\otimes n}\neq\rho_{\mathcal{E}%
_{1}}^{\otimes n}):=\frac{1-D(\rho_{\mathcal{E}_{0}}^{\otimes n}%
,\rho_{\mathcal{E}_{1}}^{\otimes n})}{2}, \label{Dstarttt}%
\end{equation}
with suitable asymptotic formulation for bosonic channels. Therefore, we may
apply Eq.~(\ref{pbbounds}) and write%
\begin{equation}
\frac{1-\sqrt{1-F(\rho_{\mathcal{E}_{0}},\rho_{\mathcal{E}_{1}})^{2n}}}{2}\leq
p_{\mathrm{err}}\leq\frac{F(\rho_{\mathcal{E}_{0}},\rho_{\mathcal{E}_{1}}%
)^{n}}{2}, \label{compp1}%
\end{equation}
where the fidelity is intended to be an asymptotic functional $F(\rho
_{\mathcal{E}_{0}},\rho_{\mathcal{E}_{1}}):=\lim_{\mu}F(\rho_{\mathcal{E}_{0}%
}^{\mu},\rho_{\mathcal{E}_{1}}^{\mu})$ for bosonic channels.

An alternate lower bound for $p(\rho\neq\sigma)$ comes from the quantum
Pinsker's inequality~\cite{Pinskerapp,Liebapp}. For any two quantum states,
$\rho$ and $\sigma$, we have
\begin{equation}
D(\rho,\sigma)\leq\sqrt{(\ln\sqrt{2})\min\{S(\rho||\sigma),S(\sigma||\rho)\}},
\end{equation}
where $S(\rho||\sigma):=\mathrm{Tr}[\rho(\log_{2}\rho-\log_{2}\sigma)]$ is the
quantum relative entropy. Consider now $\rho=\rho_{\mathcal{E}_{0}}^{\otimes
n}$\ and $\sigma=\rho_{\mathcal{E}_{1}}^{\otimes n}$. By using the additivity
of the relative entropy over tensor-product states, we may write the following%
\begin{equation}
D(\rho_{\mathcal{E}_{0}}^{\otimes n},\rho_{\mathcal{E}_{1}}^{\otimes n}%
)\leq\sqrt{n(\ln\sqrt{2})\min\{S(\rho_{\mathcal{E}_{0}}||\rho_{\mathcal{E}%
_{1}}),S(\rho_{\mathcal{E}_{1}}||\rho_{\mathcal{E}_{0}})\}}:=\sqrt{nS},
\label{lat5}%
\end{equation}
where the various functionals are asymptotic for bosonic channels, so that
$S(\rho_{\mathcal{E}_{0}}||\rho_{\mathcal{E}_{1}}):=\lim_{\mu}S(\rho
_{\mathcal{E}_{0}}^{\mu}||\rho_{\mathcal{E}_{1}}^{\mu})$. Replacing
Eq.~(\ref{lat5}) in Eq.~(\ref{Dstarttt}), we get%
\begin{equation}
p_{\mathrm{err}}\geq\frac{1-\sqrt{nS}}{2}. \label{compp2}%
\end{equation}

There is no general relation between this lower bound and the fidelity one in
Eq.~(\ref{compp1}), so that we take the optimum between them as in Eq.~(14) of
the main text. The quantum Pinsker's lower bound has in fact a different
scaling in the number of copies $n$\ and may be useful at low values of $n$.
Note that for depolarizing (or dephasing or erasure) channels, $\mathcal{E}%
_{0}$\ and $\mathcal{E}_{1}$, with probabilities $p$ and $q$, it is very easy
to compute the relative entropy between their Choi matrices. In fact, we have%
\begin{equation}
S(\rho_{\mathcal{E}_{0}}||\rho_{\mathcal{E}_{1}})=(1-p)\log_{2}{\left(
\frac{1-p}{1-q}\right)  }+p\log_{2}{\left(  \frac{p}{q}\right)  }.
\end{equation}
For bosonic Gaussian channels, one computes $S(\rho_{\mathcal{E}_{0}}^{\mu
}||\rho_{\mathcal{E}_{1}}^{\mu})$ using the formula of Ref.~\cite{PLOBapp} and
takes the limit.

An important upper bound is the quantum Chernoff bound (QCB)~\cite{QCB1app}.
For two states $\rho$ and $\sigma$, we may write%
\begin{equation}
p(\rho^{\otimes n}\neq\sigma^{\otimes n})\leq\frac{Q(\rho,\sigma)^{n}}%
{2},~~Q(\rho,\sigma):=\inf_{s\in\lbrack0,1]}Q_{s}(\rho,\sigma),~~Q_{s}%
(\rho,\sigma):=\mathrm{Tr}(\rho^{s}\sigma^{1-s}), \label{QCBppp}%
\end{equation}
satisfying the inequalities
\begin{equation}
Q(\rho,\sigma)\leq Q_{1/2}(\rho,\sigma):=\mathrm{Tr}(\rho^{1/2}\sigma
^{1/2})\leq F(\rho,\sigma)~. \label{battaetal}%
\end{equation}
Set $\rho=\rho_{\mathcal{E}_{0}}$\ and $\sigma=\rho_{\mathcal{E}_{1}}$. From
Eqs.~(\ref{Dstarttt}) and~(\ref{QCBppp}), we get%
\begin{equation}
p_{\mathrm{err}}\leq\frac{Q(\rho_{\mathcal{E}_{0}},\rho_{\mathcal{E}_{1}}%
)^{n}}{2},
\end{equation}
where $Q(\rho_{\mathcal{E}_{0}},\rho_{\mathcal{E}_{1}}):=\lim_{\mu}%
Q(\rho_{\mathcal{E}_{0}}^{\mu},\rho_{\mathcal{E}_{1}}^{\mu})$ for bosonic
channels. The QCB is asymptotically tight~\cite{QCB1app,QCB2app}, so that we
may write $p_{\mathrm{err}}\simeq Q(\rho_{\mathcal{E}_{0}},\rho_{\mathcal{E}%
_{1}})^{n}/2$ for large $n$. In many cases the QCB is easy to compute. For
example, consider qudit Pauli channels, $\mathcal{E}_{\boldsymbol{p}}$ and
$\mathcal{E}_{\boldsymbol{q}}$, with associated probability distributions
$\mathbf{p}=\{p_{k}\}$ and $\boldsymbol{q}=\{p_{k}\}$. The channels' Choi
matrices are Bell-diagonal states, so that they commute and their QCB is just%
\begin{equation}
Q(\rho_{\mathcal{E}_{\boldsymbol{p}}},\rho_{\mathcal{E}_{\boldsymbol{q}}%
})=\inf_{s\in\lbrack0,1]}\sum_{k}p_{k}^{s}q_{k}^{1-s}.
\end{equation}

\section{General connection between quantum parameter estimation and
infinitesimal quantum hypothesis testing}

We may draw a simple connection between the performance of parameter
estimation and that of infinitesimal state/channel discrimination. Consider
two equiprobable infinitesimally-close states $\rho_{_{\theta}}$ and
$\rho_{_{\theta+d\theta}}$. The $n$-copy minimum average error probability is
given by the Helstrom bound
\begin{equation}
p(\theta,n):=p(\rho_{_{\theta}}^{\otimes n}\neq\rho_{_{\theta+d\theta}%
}^{\otimes n})=\frac{1-D\left(  \rho_{_{\theta}}^{\otimes n},\rho
_{_{\theta+d\theta}}^{\otimes n}\right)  }{2}. \label{averagePP}%
\end{equation}
This probability satisfies the Fuchs-van der Graaf relations of
Eq.~(\ref{pbbounds}) with $\rho=\rho_{_{\theta}}$ and $\sigma=\rho
_{_{\theta+d\theta}}$, i.e., we may write%
\begin{equation}
\frac{1-\sqrt{1-F_{\theta}^{2n}}}{2}\leq p(\theta,n)\leq\frac{F_{\theta}^{n}%
}{2},~~~~F_{\theta}:=F(\rho_{_{\theta}},\rho_{_{\theta+d\theta}}). \label{uu2}%
\end{equation}
Now the optimal estimation of parameter $\theta$ is specified by the QCRB
\begin{equation}
V_{\theta}:=\mathrm{Var}(\theta)\geq(nI_{\theta})^{-1}, \label{QCRBmm}%
\end{equation}
where $I_{\theta}$ is the QFI, satisfying%
\begin{equation}
I_{\theta}=\frac{8(1-F_{\theta})}{d\theta^{2}},~~F_{\theta}\simeq
1-\frac{I_{\theta}d\theta^{2}}{8}+O(d\theta^{4}). \label{uu1}%
\end{equation}

Note that, at the leading order in $d\theta$, we may expand
\begin{equation}
F_{\theta}^{n}\simeq\left(  1-\frac{I_{\theta}d\theta^{2}}{8}\right)
^{n}\simeq1-\frac{nI_{\theta}d\theta^{2}}{8}\simeq\exp\left(  -\frac
{nI_{\theta}d\theta^{2}}{8}\right)  .
\end{equation}
By using the latter in Eq.~(\ref{uu2}), we get%
\begin{equation}
\frac{1-\sqrt{1-e^{-nI_{\theta}d\theta^{2}/4}}}{2}\leq p(\theta,n)\leq\frac
{1}{2}\exp\left(  -\frac{nI_{\theta}d\theta^{2}}{8}\right)  \leq\frac{1}%
{2}\exp\left(  -\frac{d\theta^{2}}{8V_{\theta}}\right)  , \label{mainuuj}%
\end{equation}
where we also use Eq.~(\ref{QCRBmm}). This equation connects the infinitesimal
error probability with the QFI and the QCRB. In particular, for large $n$, the
QCRB\ is achievable, i.e., $V_{\theta}\simeq(nI_{\theta})^{-1}$. Therefore,
for large $n$, we may write%
\begin{equation}
\frac{1-\sqrt{1-e^{-d\theta^{2}/(4V_{\theta})}}}{2}\leq p(\theta,n)\leq
\frac{1}{2}\exp\left(  -\frac{d\theta^{2}}{8V_{\theta}}\right)  .
\end{equation}

It is interesting to ask when $p(\theta,n)$ can approach the upper bound in
Eq.~(\ref{mainuuj}). This may happen when the two infinitesimally-close
quantum states $\rho_{_{\theta}}$ and $\rho_{_{\theta+d\theta}}$ are such that
the computation of their QCB reduces to the Bures fidelity, i.e.,
\begin{equation}
Q_{_{\theta}}:=\inf_{s\in\lbrack0,1]}\mathrm{Tr}(\rho_{_{\theta}}^{s}%
\rho_{_{\theta+d\theta}}^{1-s})=F_{\theta}. \label{ooopp}%
\end{equation}
Note that the latter condition is certainly satisfied if one of the two states
is pure (or both). It is also valid if: (i)~the QCB is optimal for $s=1/2$,
therefore coinciding with the quantum Battacharyya bound $Q_{1/2}$; and~(ii)
the two states commute, so that $Q_{1/2}=F$ [see Eq.~(\ref{battaetal})]. If
Eq.~(\ref{ooopp}) holds, then we may write the following asymptotic formula
for large $n$%
\begin{equation}
p(\theta,n)\simeq\frac{Q_{_{\theta}}^{n}}{2}=\frac{F_{\theta}^{n}}{2}%
\simeq\frac{1}{2}\exp\left(  -\frac{d\theta^{2}}{8V_{\theta}}\right)  ~.
\end{equation}
Let us express all these results compactly in a lemma.

\begin{lemma}
\label{lemmaAPP}Consider two infinitesimally-close quantum states,
$\rho_{_{\theta}}$ and $\rho_{_{\theta+d\theta}}$. The $n$-copy error
probability $p(\theta,n)$ defined in Eq.~(\ref{averagePP}) is bounded by the
QFI $I_{\theta}$ and the QCRB $V_{\theta}\geq(nI_{\theta})^{-1}$ as in
Eq.~(\ref{mainuuj}). In particular, if the QCB $Q_{_{\theta}}$ computed on
these states reduces to their Bures fidelity $F_{\theta}$ as in
Eq.~(\ref{ooopp}). Then, for large $n$, the error probability follow the
exponential law%
\begin{equation}
p(\theta,n)\simeq\frac{1}{2}\exp\left(  -\frac{d\theta^{2}}{8V_{\theta}%
}\right)  ,~~V_{\theta}\simeq(nI_{\theta})^{-1}.
\end{equation}

\end{lemma}

The previous lemma shows how parameter estimation bounds the performance of
infinitesimal state discrimination. We may also derive an opposite argument,
i.e., write simple inequalities showing how infinitesimal state discrimination
bounds the performance of parameter estimation. In fact, the Fuchs-van der
Graaf relations may also be inverted into the following%
\begin{equation}
2p_{\theta}\leq F_{\theta}\leq\sqrt{1-(1-2p_{\theta})^{2}},
\end{equation}
where $p_{\theta}:=p(\theta,1)=p(\rho_{_{\theta}}\neq\rho_{_{\theta+d\theta}%
})$. From this, one may easily derive%
\begin{equation}
\frac{8\left\{  1-\sqrt{1-(1-2p_{\theta})^{2}}\right\}  }{d\theta^{2}}\leq
I_{\theta}\leq\frac{8(1-2p_{\theta})}{d\theta^{2}}~.
\end{equation}
For instance, if discrimination is random ($p\rightarrow1/2$) then $I_{\theta
}\rightarrow0$, so that the QCRB tends to infinity. If the discrimination is
perfect ($p\rightarrow0$) then the QFI is unbounded $I_{\theta}\rightarrow
+\infty$, so that the QCRB tends to zero.

Note that the reasonings in this section, on the connection between parameter
estimation and infinitesimal state/channel discrimination, are not limited to
discrete-variable systems but also apply to continuous-variable (bosonic)
systems, as long as we consider asymptotic formulations for the functionals involved.

\section{Adaptive error probability for bosonic Gaussian channels}

\subsection*{Thermal-loss and amplifier channels}

Consider two thermal-loss channels, $\mathcal{E}_{0}$ and $\mathcal{E}_{1}$,
with the same transmissivity $0<\eta<1$ but different thermal noise $\bar
{n}_{0}$ and $\bar{n}_{1}$. Their asymptotic Choi matrices $\rho
_{\mathcal{E}_{0}}:=\rho(\eta,\bar{n}_{0})$ and $\rho_{\mathcal{E}_{1}}%
:=\rho(\eta,\bar{n}_{1})$ are defined by taking the $\mu$-limit over
finite-energy versions, $\rho_{\mathcal{E}_{0}}^{\mu}$ and $\rho
_{\mathcal{E}_{1}}^{\mu}$, associated with a TMSV state $\Phi^{\mu}$ at the
input. It is easy to compute their asymptotic fidelity~\cite{Banchiapp}
\begin{equation}
F(\bar{n}_{0},\bar{n}_{1})=\frac{\sqrt{2\bar{n}_{0}\bar{n}_{1}+\bar{n}%
_{0}+\bar{n}_{1}+1+2\sqrt{\bar{n}_{0}\bar{n}_{1}(\bar{n}_{0}+1)(\bar{n}%
_{1}+1)}}}{\bar{n}_{0}+\bar{n}_{1}+1}.
\end{equation}
This expression provides lower and upper bounds for the adaptive error
probability $p_{\mathrm{err}}(\bar{n}_{0},\bar{n}_{1})$ according to
Eq.~(\ref{compp1}). (It is also easy to check that one retrieves the
already-computed QFI by taking $\bar{n}_{0}=\bar{n}_{T}$ and $\bar{n}_{1}%
=\bar{n}_{T}+d\bar{n}_{T}$ and expanding at the second order.)

Let us compute the asymptotic QCB. We first compute the finite-energy QCB\ for
$\rho_{\mathcal{E}_{0}}^{\mu}:=\rho^{\mu}(\eta,\bar{n}_{0})$ and
$\rho_{\mathcal{E}_{1}}^{\mu}:=\rho^{\mu}(\eta,\bar{n}_{1})$ by using the
formula for multi-mode Gaussian states given in Ref.~\cite{QCB3app}. Then, we
take the limit for large $\mu$\ and we derive the asymptotic functional
associated with the asymptotic Choi matrices. We find
\begin{equation}
Q(\bar{n}_{0},\bar{n}_{1})=\inf_{s\in\lbrack0,1]}\left[  (\bar{n}_{0}%
+1)^{s}(\bar{n}_{1}+1)^{1-s}-\bar{n}_{0}^{s}\bar{n}_{1}^{1-s}\right]  ^{-1}.
\label{Qn0n1}%
\end{equation}
Therefore, for large $n$, the adaptive error probability scales as%
\begin{equation}
p_{\mathrm{err}}(\bar{n}_{0},\bar{n}_{1})\simeq\frac{1}{2}Q(\bar{n}_{0}%
,\bar{n}_{1})^{n}=\frac{1}{2}\inf_{s}\left[  (\bar{n}_{0}+1)^{s}(\bar{n}%
_{1}+1)^{1-s}-\bar{n}_{0}^{s}\bar{n}_{1}^{1-s}\right]  ^{-n}.
\end{equation}
We find the same results for two amplifier channels with the same gain
$\eta>1$ but different thermal noise.

As a specific example, consider two thermal-loss channels (or amplifier
channels) with infinitesimally-close thermal numbers $\bar{n}_{0}=\bar{n}%
_{T}>0$ and $\bar{n}_{1}=\bar{n}_{T}+d\bar{n}_{T}$. The minimum error
probability affecting their adaptive discrimination is%
\begin{equation}
p_{\mathrm{err}}(d\bar{n}_{T}):=p_{\mathrm{err}}(\bar{n}_{T},\bar{n}_{T}%
+d\bar{n}_{T})=\frac{1-\lim_{\mu}D^{\mu}}{2},~~~D^{\mu}:=D[\rho^{\mu}%
(\eta,\bar{n}_{T})^{\otimes n},\rho^{\mu}(\eta,\bar{n}_{T}+d\bar{n}%
_{T})^{\otimes n}],
\end{equation}
where the latter is the trace distance computed on finite-energy
Choi-approximating states. In this case, for any $\bar{n}_{T}>0$, we find that
the QCB is achieved for $s=1/2$ and that the asymptotic states $\rho(\eta
,\bar{n}_{T}):=\lim_{\mu}\rho^{\mu}(\eta,\bar{n}_{T})$ and $\rho(\eta,\bar
{n}_{T}+d\bar{n}_{T}):=\lim_{\mu}\rho^{\mu}(\eta,\bar{n}_{T}+d\bar{n}_{T})$
commute (this can be checked by diagonalizing the finite-energy versions, and
then verifying that the diagonalizing Gaussian unitaries are equal for
$\mu\rightarrow+\infty$). This means that we may write%
\begin{equation}
Q(\bar{n}_{T},\bar{n}_{T}+d\bar{n}_{T})=F(\bar{n}_{T},\bar{n}_{T}+d\bar{n}%
_{T})\simeq1-\frac{d\bar{n}_{T}^{2}}{8\bar{n}_{T}(\bar{n}_{T}+1)},
\end{equation}
which may be equivalently found by directly expanding Eq.~(\ref{Qn0n1}) at the
second order in $d\bar{n}_{T}$. Therefore, for large $n$, we derive
\begin{equation}
p_{\mathrm{err}}(d\bar{n}_{T})\simeq\frac{1}{2}Q(\bar{n}_{T},\bar{n}_{T}%
+d\bar{n}_{T})^{n}\simeq\frac{1}{2}\exp\left[  -\frac{n~d\bar{n}_{T}^{2}%
}{8\bar{n}_{T}(\bar{n}_{T}+1)}\right]  .
\end{equation}

The latter result may be equivalently derived by exploiting the connection
with parameter estimation. In fact, we can apply Lemma~\ref{lemmaAPP} with
$I_{\theta}$ given by$\ B(\bar{n}_{T}):=\lim_{\mu}B(\bar{n}_{T},\mu)=[\bar
{n}_{T}(\bar{n}_{T}+1)]^{-1}$ as in Eq.~(\ref{QFIooo}), so that%
\begin{equation}
p_{\mathrm{err}}(d\bar{n}_{T})\leq\frac{1}{2}\exp\left[  -\frac{n~d\bar{n}%
_{T}^{2}}{8\bar{n}_{T}(\bar{n}_{T}+1)}\right]  \leq\frac{1}{2}\exp\left(
-\frac{d\bar{n}_{T}^{2}}{8V_{\bar{n}_{T}}}\right)  , \label{Perrpopp}%
\end{equation}
where $V_{\bar{n}_{T}}:=\mathrm{Var}(\bar{n}_{T})\geq\lbrack nB(\bar{n}%
_{T})]^{-1}$ is the QCRB for the adaptive estimation of the thermal noise
$\bar{n}_{T}$. For large $n$, we may finally write
\begin{equation}
p_{\mathrm{err}}(d\bar{n}_{T})\simeq\frac{1}{2}\exp\left(  -\frac{d\bar{n}%
_{T}^{2}}{8V_{\bar{n}_{T}}}\right)  .
\end{equation}

\subsection*{Discriminating thermal from vacuum noise}

It is known that the computation of the fidelity, QFI and QCB may face
discontinuities at border points. For instance, see the discussions in
Refs.~\cite{Banchiapp,Safra} for the fidelity/QFI and those in
Ref.~\cite{gaeFID} for\ the fidelity/QCB. In particular, as discussed in
Ref.~\cite[Section~3]{gaeFID}, the infimum in QCB $Q:=\inf_{s}Q_{s}$\ can
always be restricted to the open interval $s\in(0,1)$. In fact, we always have
$Q=1$ at the border points $s=0,1$ and there are important cases where the
infimum is taken by the limits $s\rightarrow0^{+}$ or $s\rightarrow1^{-}$.
This is the situation when we study the discrimination of a lossy channel
($\bar{n}_{0}=0$) from an infinitesimal thermal-loss channel ($\bar{n}%
_{1}=d\bar{n}_{T}$). By replacing $\bar{n}_{T}=0$ and $\bar{n}_{1}=d\bar
{n}_{T}$ in Eq.~(\ref{Qn0n1}) and optimizing over the open interval $(0,1)$,
we find%
\begin{equation}
Q(0,d\bar{n}_{T})=\inf_{s\in(0,1)}(d\bar{n}_{T}+1)^{s-1}=\lim_{s\rightarrow
0^{+}}(d\bar{n}_{T}+1)^{s-1}=\frac{1}{d\bar{n}_{T}+1}\simeq1-d\bar{n}_{T}~,
\label{QCBnearr}%
\end{equation}
where the approximation is obtained by expanding at the first order in
$d\bar{n}_{T}\simeq0$. From Eq.~(\ref{QCBnearr}), we finally derive the
following bound for the minimum error probability affecting the adaptive
discrimination of vacuum and infinitesimal thermal noise
\begin{equation}
p_{\mathrm{err}}(d\bar{n}_{T})\leq\frac{\exp(-n~d\bar{n}_{T})}{2},
\end{equation}
which is achievable for large $n$. Note that this is different from
Eq.~(\ref{Perrpopp}) which is valid for $\bar{n}_{T}>0$.

\subsection*{Additive-noise Gaussian channels}

Consider now two additive-noise Gaussian channels, $\mathcal{E}_{0}$ and
$\mathcal{E}_{1}$, with different noise variances $w_{0}>0$ and $w_{1}>0$. For
their asymptotic Choi matrices $\rho_{\mathcal{E}_{0}}:=\rho(w_{0})$ and
$\rho_{\mathcal{E}_{1}}:=\rho(w_{1})$, we compute the asymptotic fidelity and
QCB
\begin{equation}
F(w_{0},w_{1})=\frac{2\sqrt{w_{0}w_{1}}}{w_{0}+w_{1}},~~Q(w_{0},w_{1}%
)=\inf_{s}\frac{w_{0}^{1-s}w_{1}^{s}}{(1-s)w_{0}+sw_{1}}.
\end{equation}
These quantities can be used to build lower and upper bounds for the adaptive
error probability $p_{\mathrm{err}}(w_{0},w_{1})$ affecting their
discrimination, according to Eq.~(14) of the main text. Consider now the
infinitesimal discrimination problem, setting $w_{0}=w$ and $w_{1}=w+dw$. We
find that the QCB takes the optimum at $s=1/2$ and its expansion concides with
that of the fidelity, i.e.,
\begin{equation}
Q(w,w+dw)\simeq1-\frac{dw^{2}}{8w^{2}}.
\end{equation}
From Lemma~\ref{lemmaAPP}, we derive that the adaptive error probability
$p_{\mathrm{err}}(dw):=p_{\mathrm{err}}(w,w+dw)$ satisfies%
\begin{equation}
p_{\mathrm{err}}(dw)\leq\frac{1}{2}\exp\left(  -\frac{n~dw^{2}}{8w^{2}%
}\right)  ,
\end{equation}
which is achievable for large $n$.

\section{Further remarks}

\subsection*{Relations with previous literature}

Teleportation simulation of Pauli channels was originally introduced by
Ref.~\cite{B2app}. Very recenly, this idea was generalized to any channel at
any dimension (finite or infinite) in Ref.~\cite{PLOBapp}, where channel
simulation may be realized not only by generalized teleportation protocols but
also adopting arbitrary LOCCs (with suitable asymptotic formulations for
bosonic channels). In particular, Ref.~\cite{PLOBapp} showed that the property
of teleportation covariance implies that a quantum channel can be simulated by
teleporting the input states by using the channel's Choi matrix as a resource.
This was proven at any dimension, therefore assuming asymptotic Choi matrices
for bosonic states. Previously, teleportation covariance was also considered
in Ref.~\cite{Leungapp} but restrictively to the case of discrete-variable
channels. Ref.~\cite{PLOBapp} then designed a dimension-independent technique
dubbed \textquotedblleft teleportation stretching\textquotedblright. This
technique exploits the LOCC simulation of a quantum channel to reduce an
arbitrary protocol for quantum/private communication to a much simpler block
form. Combining this reduction with the use of LOCC-contractive functionals
(such as the relative entropy of entanglement), Ref.~\cite{PLOBapp} reduced
the computation of two-way assisted quantum/private capacities to
single-letter quantities. In terms of methodology, our Letter explicitly shows
how to extend the reduction method of teleportation stretching to the realm of
quantum metrology and quantum hypothesis testing.

The quantum simulation of a channel by means of a joint trace-preserving QO
$\mathcal{U}$\ and a programme state $\sigma_{\mathcal{E}}$ traces back to the
notion of programmable quantum gate array~\cite{Gatearrayapp}. The original
idea considered the probabilistic simulation of an arbitrary unitary, but the
concept can be suitably adapted to considering the deterministic simulation of
a class of \textquotedblleft programmable\textquotedblright\ quantum channels.
This tool was considered in Refs.~\cite{Qsim0app,Qsimapp} in the context of
non-adaptive quantum metrology. Later, Ref.~\cite{ada4bisapp} realized that
its applicability can be extended to adaptive protocols. As explained in the
main text, one of the contributions of our Letter is to give a specific and
powerful design to this tool, so that $\mathcal{U}$ reduces to teleportation
and the difficult-to-find programme state $\sigma_{\mathcal{E}}$ is just the
channel's Choi matrix. This insight may potentially reduce the class of
channels but remarkably simplifies computations. Furthermore, it allows us to
establish a simple \textquotedblleft golden rule\textquotedblright%
\ (teleportation covariance) for the identification of channels that are
simulable by teleportation and, therefore, programmable. As also discussed in
the main text, the reduction to the channel's Choi matrix brings non-trivial advantages:

\begin{description}
\item[(1)] The generally-adaptive QFI\ is easily computable. For instance,
compare our formula [Eq.~(7) of the main text] expressing the adaptive QFI for
a teleportation-covariant channel in terms of its Choi matrix, with the more
general but much more difficult formula in Eq.~(9) of Ref.~\cite{ada4bisapp}%
\ which involves the minimization of the operator norm of sum of derivatives
of Kraus operators over different Kraus representations of the channel.
Because of this drastic simplification, we can compute the adaptive QFI for
many channels in a completely trivial way, and we may go beyond the results
previously known. For instance, we may consider arbitrary Pauli channels (not
just depolarizing/dephasing channels) for which we may also consider
multi-parameter noise estimation. Most importantly, we may extend the results
to bosonic Gaussian channels thanks to the fact that we may apply well-known
teleportation-based simulations developed for continuous-variable systems.

\item[(2)] The QCRB is asymptotically achievable without adaptiveness for any
teleportation-covariant channel. The asymptotic expression of the QCRB is
given by $[nB(\rho_{\mathcal{E}_{\theta}})]^{-1}$ where $B$\ is the QFI
computed on the channel's Choi matrix $\rho_{\mathcal{E}_{\theta}}$. For a
programmable channel with programme state $\sigma_{\mathcal{E}_{\theta}}$\ the
QCRB is bounded by $[nB(\sigma_{\mathcal{E}_{\theta}})]^{-1}$, but the latter
is not generally achievable unless $\sigma_{\mathcal{E}_{\theta}}$ can be
generated as an output from the channel. For a generic programmable channel,
it is an open problem to show that the optimal scaling is achievable without
adaptiveness. In terms of the classification of metrological schemes defined
in Ref.~\cite{ada4bisapp}, we have proven $\mathrm{(iii)}=\mathrm{(iv)}$ for
any teleportation-covariant channel (at any dimension), while one still has
$\mathrm{(iii)}\leq\mathrm{(iv)}$ for a generic programmable channel. Here
$\mathrm{(iii)}$ is an optimal entanglement-assisted protocol (with passive
ancillas, without feedback), while $\mathrm{(iv)}$ is an optimal adaptive protocol.
\end{description}

\subsection*{Main achievements of this work}

It may be useful to give a schematic list of the main achievements of our work:

\begin{description}
\item[1- Teleportation as primitive for quantum metrology (no-go theorem).]
For the first time, we establish a direct connection between teleportation and
quantum metrology. We prove a general no-go theorem that can be summarized as
follows: The ultimate estimation of noise parameters in
teleportation-covariant channels cannot beat the SQL. Furthermore we show that
the optimal scaling is achievable by just using entanglement without the need
of adaptive protocols (this is still unproven for generic programmable
channels). As already discussed before, the class of teleportation-covariant
channels is extremely wide, including discrete-variable channels such as Pauli
and erasure channels (at any finite dimension) besides continuous-variable
channels such as bosonic Gaussian channels. As a matter of fact, the
teleportation-based approach is so powerful and general that it is an open
problem to find other channels (e.g., programmable) for which we may compute
the adaptive QFI beyond the class of teleportation-covariant channels.

\item[2- Analytical formulas for adaptive noise estimation.] We compute a
number of analytical formulas for the ultimate quantum Fisher information in
adaptive noise estimation. These are remarkably simple formulas in terms of
the Choi matrices of the encoding channels. Setting the limits for estimating
decoherence and noise has broad implications, e.g., for protocols of quantum
sensing, imaging and tomography.

\item[3-~Ultimate adaptive estimation of thermal noise.] We set the ultimate
limit for estimating thermal noise in a bosonic Gaussian channel (thermal-loss
or amplifier channel). The thermal-loss channel is particularly important
because its dilation represents a basic model of eavesdropping in
continuous-variable quantum key distribution (known as \textquotedblleft
entangling cloner\textquotedblright\ attack~\cite{WeeRMPapp}). The exact
quantification of this noise is a crucial step for deciding how much error
correction and privacy amplification is needed for the practical agreement of
a secret key. These results are also extended to the additive-noise Gaussian
channel and can be used to bound the performance of adaptive measurements of
temperature (e.g., in quasi-monochromatic bosonic baths).

\item[4-~Ultimate limits of adaptive channel discrimination.] Our derivations
can be applied to other scenarios. In quantum channel discrimination, we show
that the ultimate error probability for distinguishing two
teleportation-covariant channels is uniquely determined by the trace distance
between their Choi matrices. This also means that, for these channels, optimal
strategies do not need feedback-assistance. By drawing a simple connection
between parameter estimation (quantum metrology) and discrimination (quantum
hypothesis testing), we then derive a simple formula for the ultimate
resolution of two extremely-close temperatures.
\end{description}


\begin{thebibliography}{99}
\bibitem {averageLOCC}The output state is implicitly averaged over the
outcomes of all measurements performed in the protocol.
\bibitem {BuresFID}This is the Bures' fidelity, equal to the square-root of
the Uhlmann's fidelity~\cite{Fid1,Fid2}.
\bibitem {parameterU}Note that, in general, one may allow for a weak $\theta
$-dependence in $\mathcal{U}$. For quantum parameter estimation, one can write
Eq.~(\ref{NCppp}) up to $O(d\theta^{2})$~\cite{Qsim}.
\bibitem {Achievability}Suppose that we repeat our reasonings for a generic
programmable channel $\mathcal{E}_{\theta}$ with programme state
$\sigma_{\mathcal{E}_{\theta}}$. We can modify the (finite-dimensional) proofs
and write the output state $\rho_{\mathbf{ab}}^{n}(\theta)=\tilde{\Lambda
}(\sigma_{\mathcal{E}_{\theta}}^{\otimes n})$ for a CPTP map $\tilde{\Lambda}%
$, leading to the bound $\bar{I}_{\theta}^{n}\leq nB(\sigma_{\mathcal{E}%
_{\theta}})$ (see also Ref.~\cite{ada4bis}). Unfortunately, the latter bound
is not achievable unless one shows an explicit protocol where $\sigma
_{\mathcal{E}_{\theta}}$ is generated at the channel's output. This is fully
solved by our Theorem~\ref{theoSTT}, where the Choi matrix $\rho
_{\mathcal{E}_{\theta}}$ not only \textquotedblleft
programmes\textquotedblright\ the channel but can also be generated by
propagation through $\mathcal{I}\otimes\mathcal{E}_{\theta}$, so that we may
write an equality in Eq.~(\ref{firstTH}). As a result, the QCRB scales
asymptotically as $\mathrm{Var}(\theta)\simeq\lbrack nB(\rho_{\mathcal{E}%
_{\theta}})]^{-1}$ and the optimal scaling is reached by non-adaptive
strategies. We cannot state these results for a generic programmable channel,
for which the optimal estimation strategy can still be adaptive.
\bibitem {lossnoise}In particular, this extension regards the estimation of
noise parameters (thermal background) but not loss parameters. For the latter,
we always obtain the trivial bound $\mathrm{Var}(\theta)\geq0$ which comes
from $\lim_{\mu}F(\rho_{\mathcal{E}_{\theta}}^{\mu},\rho_{\mathcal{E}%
_{\theta+d\theta}}^{\mu})=0$. In fact, we can always perfectly distinguish and
estimate two infinitesimally-close transmission parameters in the limit of
infinite input energy. Finding the optimal adaptive estimation of the loss
parameter of a Gaussian channel with an input energy constraint is an open
problem subject to investigation.
\bibitem {MethodsNOTA}The methodology can be applied to other types of
adaptive protocols. The first ingredient is the simulation of a programmable
channel~\cite{Gatearray} which is teleportation-based\ for a
teleportation-covariant channel~\cite{PLOB}. Using this simulation, we may
reduce arbitrary $n$ uses of an adaptive protocol into a block form so that
its output has decomposition $\tilde{\Lambda}(\sigma_{\mathcal{E}}^{\otimes
n})$ for a CPTP map $\tilde{\Lambda}$ and a programme state $\sigma
_{\mathcal{E}}$, the latter being the channel's Choi matrix $\rho
_{\mathcal{E}}$ for a teleportation-covariant channel. Now for any functional
$f$ monotonic/contractive under CPTP maps, we\ may discard $\tilde{\Lambda}$
and write an upper bound. Furthermore, if the functional is subadditive over
tensor products (QFI, relative entropy, trace distance, Holevo bound, quantum
mutual information) or multiplicative (fidelity, quantum Chernoff bound), then
we get a single-letter bound.
\end{thebibliography}
\end{document}